\documentclass[superscriptaddress,amsmath,amssymb,aps,twocolumn,praplied,notitlepage]{revtex4-2}
\usepackage{amsmath,amsfonts,amssymb,amsthm,graphics,graphicx,epsfig,bbm}
\usepackage[colorlinks=true,citecolor=blue,linkcolor=blue,urlcolor=blue]{hyperref}
\usepackage[usenames]{color}
\usepackage{graphicx}
\usepackage{subfigure}
\usepackage{amsmath}
\usepackage{epsfig}
\usepackage{dcolumn}
\usepackage{bm}
\usepackage{color}
\usepackage{xfrac}
\usepackage{times}
\usepackage{epstopdf}
\usepackage{amssymb}
\usepackage{amstext}
\usepackage{latexsym}
\usepackage{hyperref}
\usepackage{amsfonts}
\usepackage{psfrag}
\usepackage{soul,xcolor}
\usepackage[normalem]{ulem}
\usepackage{booktabs}
\usepackage[table]{xcolor}
\usepackage{adjustbox}

\newcommand{\ket}[1]{\vert #1 \rangle}
\newcommand{\bra}[1]{\langle #1 \vert}
\newcommand{\ketbra}[2]{\vert #1 \rangle \langle #2 \vert}

\newcommand{\abs}[1]{| #1 |}

\newcommand{\Tr}{\mathrm{Tr}}

\begin{document}

\title{Analytical blueprint for 99.999\% fidelity X-gates on present superconducting hardware under strong driving}

\author{Jos\'e Diogo Da Costa Jesus}
\affiliation{Institute for Quantum Control (PGI-8), Forschungszentrum J\"ulich, 52425 J\"ulich, Germany}
\affiliation{Institute for Theoretical Physics, University of Cologne, 50937 Cologne, Germany}
\email{j.da.costa.jesus@fz-juelich.de}
\author{Boxi Li}
\affiliation{Institute for Quantum Control (PGI-8), Forschungszentrum J\"ulich, 52425 J\"ulich, Germany}

\author{Yuan Gao}
\affiliation{Institute for Functional Quantum System (PGI-13), Forschungszentrum J\"ulich, 52425 J\"ulich, Germany}
\affiliation{Department of Physics, RWTH Aachen University, 52074 Aachen, Germany }

\author{Rami Barends}
\affiliation{Institute for Functional Quantum System (PGI-13), Forschungszentrum J\"ulich, 52425 J\"ulich, Germany}
\affiliation{Department of Physics, RWTH Aachen University, 52074 Aachen, Germany }

\author{Francisco Andr\'es C\'ardenas-L\'opez}
\affiliation{Institute for Quantum Control (PGI-8), Forschungszentrum J\"ulich, 52425 J\"ulich, Germany}

\author{Felix Motzoi}
\affiliation{Institute for Quantum Control (PGI-8), Forschungszentrum J\"ulich, 52425 J\"ulich, Germany}
\affiliation{Institute for Theoretical Physics, University of Cologne, 50937 Cologne, Germany}

\date{\today}

\begin{abstract}
Achieving ultrafast single-qubit gates that approach the limits set by decoherence requires operating in the strong-driving regime, where conventional semi-classical descriptions and single-leakage models break down, and multi-photon transitions emerge as dominant error channels that grow rapidly as the gate time is reduced. This makes it essential to develop systematic methods to characterize and suppress these errors using pulse shapes that remain simple and experimentally straightforward to calibrate. In contrast to standard DRAG approaches for single-qubit gates, which suppress  leakage within an effective three-level description, our framework systematically treats multi-photon transitions that arise in the strong-driving regime due to the participation of higher energy levels. In particular, we identify and suppress the two-photon $\ket{1}\leftrightarrow\ket{3}$ channel, incorporate time-ordering corrections, and derive a recursive analytical construction that captures the dominant higher-order error processes. We term the resulting pulse families R1D and R2D. These constructions extend and unify DRAG-type approaches by incorporating higher-order processes and time-ordering effects, yielding analytical expressions for optimal quadrature corrections, detunings, and prefactors. In addition, we clarify the optimal choice of DRAG prefactors and constant detuning when time-ordering effects are properly taken into account. This provides a systematic prescription for calibrating additional correction terms to further improve gate performance. Using this framework, we numerically demonstrate gate infidelities below $10^{-5}$ for a $7\mathrm{ns}$ full $\pi$-rotation under realistic decoherence rates.

\end{abstract}

\maketitle

\section{Introduction}
Superconducting quantum circuits~\cite{Krantz2019,Blais2021Circuit} have emerged as a leading platform for quantum computation due to their tunability and the high degree of control achievable over individual qubits. State-of-the-art quantum processors based on transmon circuits~\cite{PhysRevA.76.042319} are well described as weakly anharmonic oscillators. However, this weak anharmonicity allows the control drive to induce unwanted transitions to higher energy levels, resulting in population leakage out of the computational subspace and phase errors, which ultimately limit single-qubit gate fidelities~\cite{PhysRevLett.116.020501}.

A widely used strategy to mitigate these errors is pulse shaping via the Derivative Removal by Adiabatic Gate (DRAG) protocol~\cite{motzoi2009simple,PhysRevA.88.062318,Theis2018Counteracting}. In this approach, an additional quadrature component is introduced to suppress off-resonant single-photon transitions and associated phase errors within a perturbative, few-level description~\cite{motzoi2009simple,PhysRevA.83.012308}. However, as gate times decrease and the drive amplitude approaches the anharmonicity, the perturbative regime breaks down and the conventional three-level description becomes insufficient, with multiple leakage channels simultaneously activated, thus leading to increasingly complex dynamics that are not captured by conventional DRAG leakage modeling~\cite{Li2024Experimental,li2024universal}. While alternative approaches based on spectral shaping have been proposed to suppress specific transitions~\cite{PRXQuantum.5.030353,Chiaro2025}, these methods rely on semi-classical approximations that are not systematically valid beyond the perturbative regime.

In this work, we develop a framework that goes beyond these limitations by explicitly accounting for multi-photon processes entirely outside the computational space and the population of higher energy levels from the transmon circuit. We show that, once single-photon leakage is suppressed, two-photon transitions -- particularly the $\ket{0}\leftrightarrow\ket{2}$ the $\ket{1}\leftrightarrow\ket{3}$ channels -- become the dominant sources of error. To address this, we construct recursive DRAG \cite{Li2024Experimental} type corrections that simultaneously suppress single- and two-photon leakage processes through a sequence of frame transformations. This leads to fully analytical pulse families that eliminate the leading error terms in the strong-driving regime, i.e.~where the pulse amplitude cannot be considered small compared to the detuning. In particular, we demonstrate that a minimum four-level description is required to capture the relevant dynamics. Using these analytical pulses without additional calibration, we achieve gate errors at the level of $10^{-5}$ for an $11.8 \mathrm{ns}$ pulse. 

Beyond this analytical construction, we provide a systematic understanding of commonly used calibration parameters. While additional control prefactors are often introduced empirically to improve performance, their role and interdependence remain not yet understood ~\cite{PhysRevLett.116.020501}. Using a combination of toggling-frame transformations and a Magnus expansion, we explain the observed correlations between these parameters. In particular, we clarify the relationship between the DRAG prefactor and detuning, and show why, counterintuitively, a constant detuning can outperform time-dependent detuning in relevant regimes. By incorporating additional prefactors to the R2D pulses we reduce the gating time from $11.8 \mathrm{ns}$ to $6.8 \mathrm{ns}$ achieving gate fidelity errors of \(10^{-5}\).

The remainder of the paper is organized as follows: Sec.~\ref{DRAG} introduces the DRAG formalism and the first-order errors on single-qubit gates. Section~\ref{R1D} describes the recursive DRAG approach, termed as R1D,  for correcting two-photon errors $\ket{0}\rightleftarrows\ket{2}$. Sec.~\ref{R2D} generalizes to a second recursive set of frame transformations, termed as R2D, that further remove errors from the higher excited states of the ladder. Sec.~\ref{Numerics} provides analytical predictions and analysis for the drive amplitude and constant detuning to remove additional computational-space errors. In Sec.~\ref{dissipation}, we analyse the effect of decoherence on the gate performance. Finally, in Sec.~\ref{Conclu} we provide our conclusions.

\section{Error budget and corresponding corrections}
\label{analytical section}

\subsection{Linear leakage error and standard DRAG correction}
\label{DRAG}

\begin{figure*}[t!]
    \centering
    \includegraphics[width=0.9\linewidth]{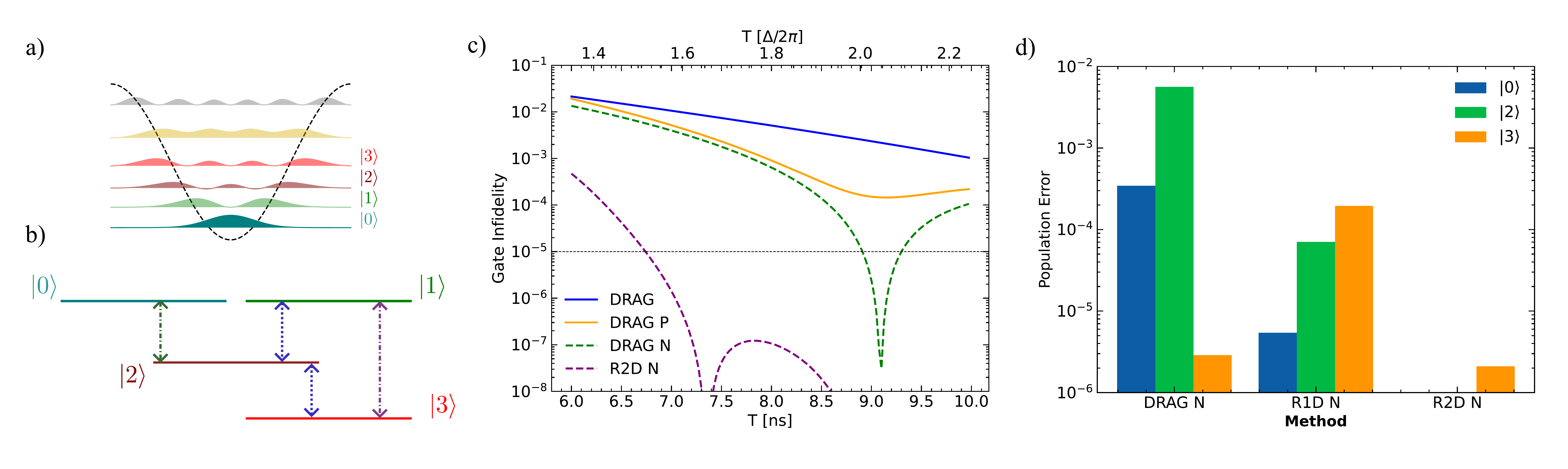}
    \caption{ \textbf{Pulse performance for a transmon qubit.} a) Low-lying energy structure of a transmon system represented as a particle in a cosine potential, where different opacity refers to the importance of higher energy levels in the performance of the gate. {Only the lowest 4 levels are considered in the analytical expressions}. b) Structure of the energy levels of the transmon system in the rotating frame with respect to the transition $\omega_{10}$. The dashed arrows connecting different states represent single-photon transitions, while dotted dashed connects states through two-photon processes. 
     c) {Gate infidelity as a function of gate time for analytical linear DRAG (blue, \ref{SupraLinear}), linear DRAG using the predictions for $\alpha$, $\beta$ and $\delta_c$ (DRAG P, orange, \ref{alpha_pred}), linear DRAG with optimized parameters (DRAG N, dashed green, \ref{drag_num}) and R2D with optimized parameters, $\alpha_{12}$, $\alpha_{02}$, $\alpha_{13}$,  $\beta$ and $\delta_c$  (R2D N, dashed purple, \ref{r2d_num}).  
    The parameterized R2D pulse outperforms all others for all gate times considered, achieving a fidelity of 99.999\%  at 6.75 ns and beyond.} d) Histogram with the final population of the relevant leakage states for  {different calibrated pulses when implementing an X gate at 7~ns. The main source of error in DRAG N, \ref{drag_num}, can be addressed by using R1D N, \ref{r1d_num}, which will suppress 2 photon transitions to $|2\rangle$. However, for R1D N the leakage to $|3\rangle$ is a major source of error. This can suppressed by more than an order of magnitude by R2D N,  \ref{r2d_num})}. The simulations were performed by modeling the transmon as an anharmonic oscillator with anharmonicity $\Delta_{2} = -2\pi \times 0.225$ GHz {and $\Delta_{3}= 3\Delta_2$}.}
    \label{fig:fig1_inks}
\end{figure*}

We start by revisiting the DRAG method introduced in Ref.~\cite{motzoi2009simple}, but considering a weakly anharmonic oscillator with four levels, $|0\rangle$, $|1\rangle$, $|2\rangle$ and $|3\rangle$, as depicted in Fig.~\ref{fig:fig1_inks}(a). The first two levels are used for information processing -- they form the computational subspace -- with the rest representing leakage states. DRAG aims to implement operations between the lowest two levels while suppressing leakage and phase errors by engineering the drive envelope. Let us start with our system Hamiltonian in the rotating frame with respect to the drive frequency, taking $\hbar =1$ (see Appendix~\ref{derivationH} for a full derivation):
\begin{equation}
    \hat{\mathcal{H}}(t) = \sum_{j=1}^M \left[\left(\Delta_j+j\delta(t)\right)|j\rangle \langle j|
    +\frac{\lambda_j}{2}\sum_{p=x,y}\Omega_p(t) \hat{\sigma}^p_{j-1, j}\right]
    \label{Exp:RWA}
\end{equation}

where $\Delta_{j}$ is the anharmonicity of the $j$th level, $\delta(t)$ is a time-dependent detuning to be determined later and $\lambda_j$ denotes the relative drive strength between $\ket{j}\rightleftarrows\ket{j+1}$. Without loss of generality, we assume $\lambda_1=1$. For transmon circuits~\cite{PhysRevA.76.042319, Khani2009Optimal}, we can approximate $\lambda_j\equiv\sqrt{j}$. The symbols $\Omega_{x}(t)$ [$\Omega_{y}(t)$] are the in-phase (out-of-phase) drive envelopes. Finally $\hat{\sigma}^x_{j,k}= |j\rangle \langle k| + |k\rangle \langle j|$ and  $\sigma^y_{j,k}= i(|j\rangle \langle k| - |k\rangle \langle j|)$ are the generalized Pauli operators acting on multi-level systems.

The goal is to implement an on-resonant interaction between the lowest two levels (\(\Delta_1=0\)) while avoiding any population transfer to higher levels.
DRAG achieves this by moving to an effective frame in which the ancillary levels decouple from the computational subspace, and by determining the corresponding corrections to the drive envelopes. The Hamiltonian in this frame reads $\hat{\mathcal{H}}_{1}(t) = \hat{V}(t)\hat{\mathcal{H}}(t)\hat{V}^{\dag}(t)+i \dot{\hat{V}}(t)\hat{V}^{\dag}(t)$, where overdot represents derivative with respect to time. 
Ideally, the frame transformation \(\hat{V}\) is chosen such that \(\hat{V}(t)\hat{\mathcal{H}}(t)\hat{V}^{\dag}(t)\) becomes block diagonal, eliminating couplings between the computational subspace and the ancillary levels~\cite{Li2022Nonperturbative}.

Because computing the exact \(\hat{V}\) is typically expensive, a perturbative Schrieffer–Wolff approach~\cite{PhysRev.149.491} is commonly used, though in some cases non-perturbative transformations are possible \cite{PhysRevA.88.062318,Li2024Experimental, ibanez2012multiple, PhysRev.149.491}.
Let $\hat{S}_{1}(t)$ denote the generator of $\hat{V}(t)$, $\hat{V}(t)= \exp({\hat{S}_{1}(t)})$. The transformed Hamiltonian to second order is 
\begin{eqnarray}\nonumber
\label{heff1}
\hat{\mathcal{H}}_{1}(t) &\approx& \hat{\mathcal{H}}(t) + [\hat{S}_{1}(t),\hat{\mathcal{H}}(t)]\\
&+& \frac{1}{2!}[\hat{S}_{1}(t),[\hat{S}_{1}(t),\hat{\mathcal{H}}(t)]] + i \dot{\hat{S}}_{1}(t).  
\end{eqnarray}
The generator $\hat{S}_{1}(t)$ generally corresponds to an anti-hermitian version of the interaction that we want to remove~\cite{Li2022Nonperturbative}. For the first order it is
\begin{eqnarray}
    \hat{S}_{1}(t) = -i\sum_{j=1}^{M}  \lambda_j \frac{\Omega_x(t)}{2\Delta_2} \hat{\sigma}^y_{j-1, j}.
    \label{eq:frame_DRAG}
\end{eqnarray}
Substituting this into Eq.~(\ref{heff1}), expanding to second order in $\Omega_{x}/\Delta_2$, {truncating to the lowest four levels yields:}
\begin{eqnarray}
\label{FirstH}
 \hat{\mathcal{H}}_{1}(t) &\approx& \frac{\Omega_{x}}{2}  \hat{\sigma}_{0,1}^x+ \bigg[\frac{\Omega_y}{2}+\frac{\dot{\Omega}_x}{2\Delta_2}\bigg] (\hat{\sigma}_{0,1}^y+\lambda_{2}\hat{\sigma}_{1,2}^y) \\ \nonumber
 &&+\big[\delta(t) +\mathcal{E}_{11}\Omega_x^2 \big]|1\rangle\langle 1| +  \Omega_x^2[\mathcal{E}^x_{02}\hat{\sigma}_{0,2}^x +  \mathcal{E}^x_{13}\hat{\sigma}_{1,3}^x] \\  \nonumber
  &&{+\big[\Delta_2 +2\delta(t) +\mathcal{E}_{22}\Omega_x^2 \big]|2\rangle\langle 2| +\Omega_x \mathcal{E}^x_{23}\hat{\sigma}_{2,3}^x}\\  \nonumber
  &&{+\big[\Delta_3 +3\delta(t) +\mathcal{E}_{33}\Omega_x^2 \big]|3\rangle\langle 3|}, 
\end{eqnarray}
where to avoid cumbersome notation we omit time-dependency in $\Omega$ and use general prefactors $\mathcal{E}_{ij}$ for error in the $i,j$-th matrix element. Maintaining the approximation order $O\big(\Omega_x^3/\Delta_2^2\big)$, here and throughout the rest of this section, yields
$\mathcal{E}_{11}=(4-\lambda_{2}^2)/4\Delta_2$, $\mathcal{E}^x_{02}=\lambda_{2}/8\Delta_2$, $\mathcal{E}^x_{13}=\lambda_{2}\lambda_{3}(\Delta_3-2\Delta_2) /8\Delta_2^2$, $\mathcal{E}^x_{23}=\lambda_3(2\Delta_2-\Delta_3)/2$, $\mathcal{E}_{22}=(\lambda_3^2(\Delta_2-3\Delta_3)+\Delta_2(2+\lambda_2^2))/(4\Delta_2^2)$, $\mathcal{E}_{33}=(2\Delta_2+(3\Delta_2-\Delta_3)\lambda_3^2)/(4\Delta_2^2)$.
 
Eq.~(\ref{FirstH}) is central to our analysis, as it identifies the dominant error mechanisms. The leading error arises from the linear coupling term \(\sigma_{1,2}^y\) with transition frequency offset $\Delta_{2}$ which motivates the standard DRAG correction $\Omega_{y}(t)=-\dot{\Omega}_{x}(t)/\Delta_{2}$~\cite{motzoi2009simple} as depicted in Fig.~\ref{fig:fig1_inks}(b), where we see the dominant error channels in the rotating frame with respect to the transition frequency $\omega$. The phase accumulated on $\ket{1}$ can be removed by dynamically modulating the detuning following the relation $\delta(t) =-\Omega_x^2(4-\lambda_{2}^2)/4\Delta_2$.
In practice, this is often approximated by a constant detuning shift~\cite{PhysRevA.83.012308,PhysRevLett.116.020501}, which we formally explain in subsequent sections.
Applying these corrections yields { for the computational subspace and its associated transitions}:
\begin{equation}
 \hat{\mathcal{H}}_{1}(t) =\frac{1}{2} \Omega_{x} \hat{\sigma}_{0,1}^x+ \Omega_x^2[\mathcal{E}^x_{02}\hat{\sigma}_{0,2}^x +  \mathcal{E}^x_{13}\hat{\sigma}_{1,3}^x]+\dots\nonumber 
\end{equation}
The remaining leakage channels are thus the two-photon processes \(\sigma_{0,2}^x\) and \(\sigma_{1,3}^x\). These contributions are negligible for weak drives but become significant as the gate time is reduced and $\Omega_x/\Delta_2$ increases, as illustrated in Fig. \ref{fig:fig1_inks} (d), where a significant leakage to $|2\rangle$  can be observed for a {calibrated DRAG pulse at 7ns}. In the following subsections, we focus on eliminating these higher-order leakage terms.

\subsection{Rotation angle and fidelity metric}\label{sec:trialfn}

Before proceeding, we first consider the constraints on the initial pulse shape prior to applying the DRAG correction.
In general, we require the pulse to be smooth and vanish at both its beginning and end. This leads to the conditions
\begin{subequations}
\begin{eqnarray}
\label{Exp1}
&&\int_0^T \Omega_x(t)dt = \theta, \\
\label{Exp2}
&&\Omega_x(0) =\Omega_x(T)=\dot{\Omega}_x(0) =\dot{\Omega}_x(T) =0,
\end{eqnarray}
\end{subequations}
which guarantee that the frame generated by $\hat{V}(t)$ coincides with the laboratory frame at the beginning and at the end of the dynamics.
To benchmark gate performance, we compute the average gate infidelity defined as $\mathcal{E}(\hat{\mathcal{U}}_Q)=1-\mathcal{F} [\hat{\mathcal{U}}_Q]$ where $\mathcal{F} [\hat{\mathcal{U}}_Q]$ can be found in Appendix~\ref{Fid_metric}, following closely~\cite{Pedersen2007Fidelity}.

\begin{figure}
    \centering
    \includegraphics[width=1\linewidth]{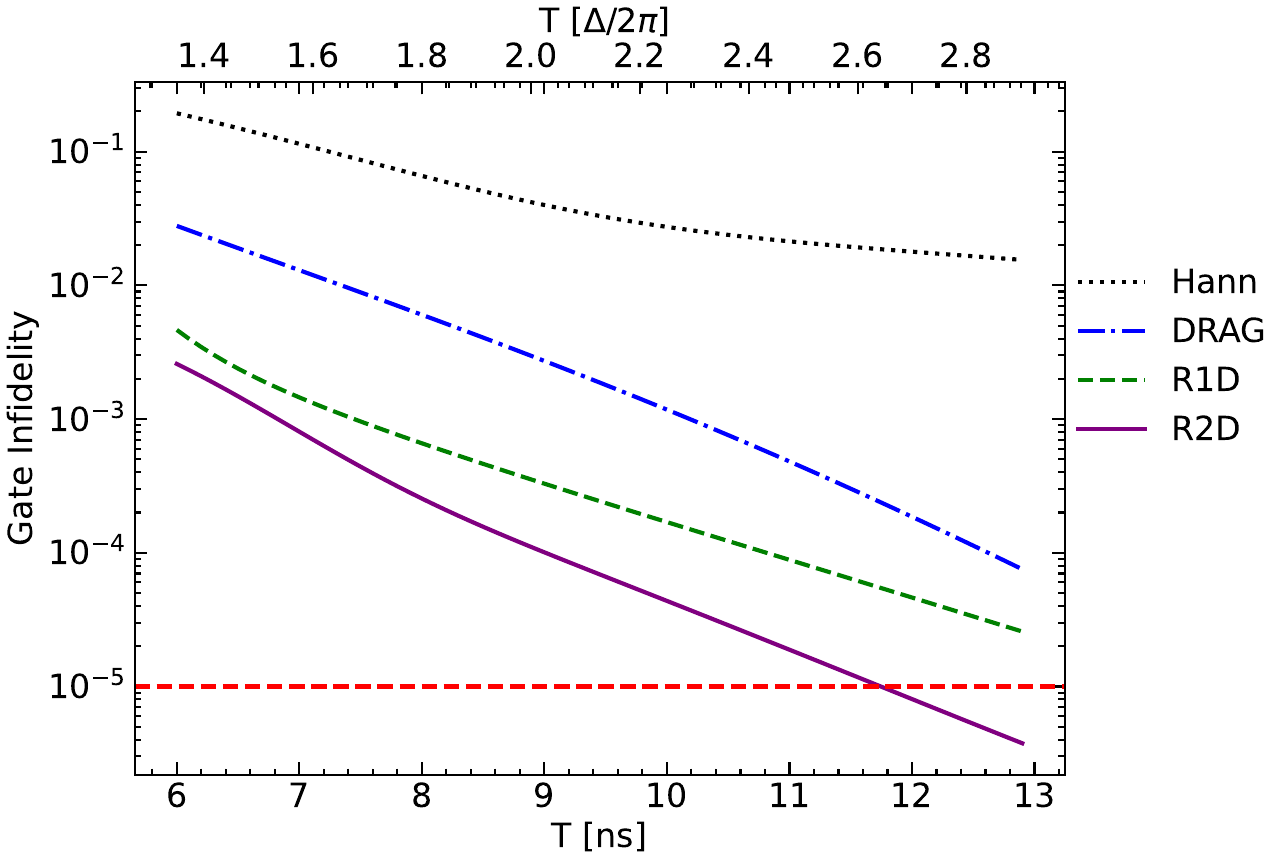}
    \caption{{\textbf{Gate infidelity for implementing an X gate under different pulses}. The black line corresponds to the Hann pulse, the blue line is for standard DRAG and green and violet lines stand for recursive pulses R1D and R2D, respectively. These pulses are simulated analytically without calibration. R2D outperforms every method and achieves a gate fidelity of 99.99 \% at 9.1~ns and a fidelity of 99.999\% at 11.8 ~ns.}}
    \label{fig:ana_fids}
\end{figure}

We compare different implementations using the Hahn pulse $\Omega_{\rm{Hann}}(t) = \Omega_0 \sin^2(\pi t/T)$ and the DRAG pulse $\Omega_{\rm{DRAG}}(t)=\Omega_{\rm{Hann}}(t)-i\dot{\Omega}_{\rm{Hann}}(t)/\Delta_{2}$ for different gate times $T$. In {Fig.~\ref{fig:ana_fids}}, we can see the gate infidelity for both protocols, for an anharmonicity of $\Delta_2/2\pi=-0.225$ GHz. This value for anharmonicity is used for all simulations throughout this paper. {Although we only considered 4 levels for the analytical derivations, all simulations were performed considering 6 levels to confirm the validity of the model considered.}

We observe that linear DRAG outperforms the Hann pulse at every gate time,  obtaining errors below $10^{-4}$ for times longer than $\approx$ 12.7~ns. However, for gate times shorter than $\approx$10~ns we see that the performance of DRAG falls sharply with fidelities below 99.9\%, demonstrating that at shorter times additional leakage channels (two-photon transitions) are the leading sources of errors. 

\begin{figure}
    \centering
    \includegraphics[width=1\linewidth]{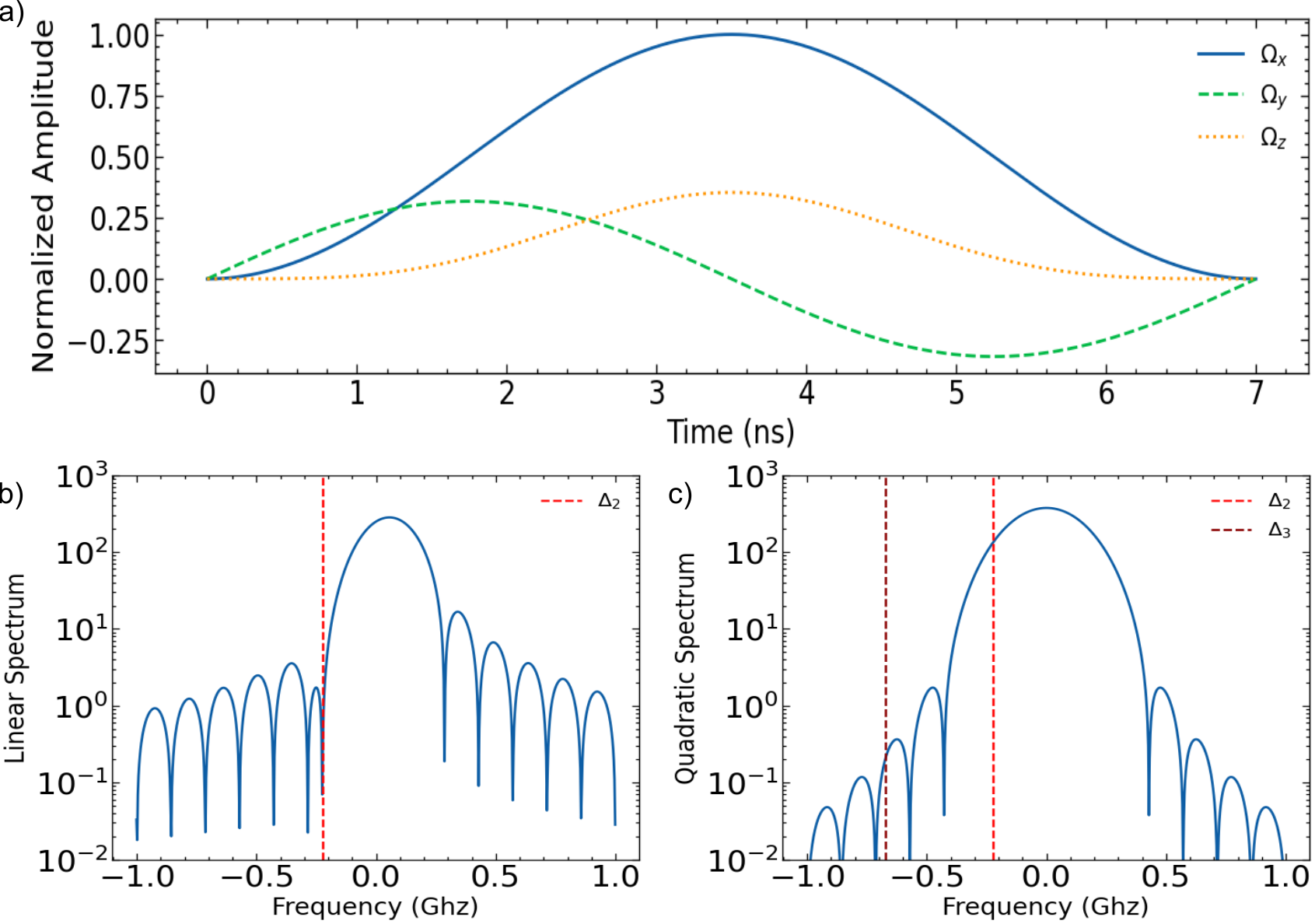}
    \caption{{\textbf{DRAG pulse shape and Fourier spectrum of $\Omega$ and $\Omega_x^2$} a) Pulse shape of a DRAG pulse starting with $\Omega_x= \Omega_0 \sin^2({2\pi t/T})$ for a gate time of $T= 7ns$. b) Fourier spectrum of a DRAG pulse, $\Omega = \Omega_x + i \Omega_y$. As can be seen, a whole in the spectrum has been added to the frequency $\Delta_2$, suppressing the $\ket{1}\rightleftarrows\ket{2}$ transition. c) Fourier spectrum of $\Omega_x^2$. As can be seen, we have significant powers at $\Delta_2$ and $\Delta_3$, the corresponding frequencies in $\Omega_x^2$ for the $\ket{0}\rightleftarrows\ket{2}$ and $\ket{1}\rightleftarrows\ket{3}$ transition, which need to be suppressed.} }
    \label{fig:drag_pulseFT}
\end{figure}
{An illustration of the pulse shape obtained using DRAG can be found in Fig. \ref{fig:drag_pulseFT} as well as the Fourier spectrum of the pulse, $\Omega= \Omega_x + i\Omega_y$, and the Fourier spectrum of the square of its real component $\Omega_x^2$. Although we have suppressed the $\ket{1}\rightleftarrows\ket{2}$ transition, as can be seen by the hole in the Fourier spectrum  of $\Omega$ at $\Delta_2$, transitions $\ket{0}\rightleftarrows\ket{2}$ and $\ket{1}\rightleftarrows\ket{3}$ are left unsupressed, as seen by the Fourier spectrum of $\Omega_x^2$.  }

\subsection{Virtual leakage $\ket{0}\rightleftarrows\ket{2}$ and R1D pulse}
\label{R1D}

As mentioned in the previous subsection, to obtain a shorter gate time without compromising on fidelity, we need to address multiple leakage pathways arising from two-photon transitions. To simultaneously mitigate these errors, we employ the DRAG method recursively~\cite{PhysRevA.88.062318,Theis2018Counteracting,Li2024Experimental}, where each successive frame transformation targets a specific leakage channel. { This method was first employed in \cite{PhysRevA.88.062318} to target the two-photon process \(\hat{\sigma}_{0,2}^x\), in a simplified 3-level ladder transmon model. Here we review this approach in the context of 4-level ladder system leakage. }

To eliminate the first two-photon process \(\hat{\sigma}_{0,2}^x\), we move to a {new} frame generated by $\hat{S}_2(t) = -i\Omega_1^2 \mathcal{E}^x_{02}\hat{\sigma}_{0,2}^y/\Delta_2$, where $\Omega_1$ is now a new trial function. {This generator was chosen as in Sec.~\ref{sec:trialfn}, i.e., to eliminate a generic transition element $\Omega \sigma^x$, we use the generator proportional to $-i\Omega \sigma^y$}. 
This transforms the Hamiltonian to 
\begin{equation}
    \hat{\mathcal{H}}_2(t) \approx\hat{\mathcal{H}}_1(t) + \mathcal{E}^x_{02} \bigg[\frac{2\Omega_1\dot\Omega_1}{\Delta_2}\hat{\sigma}_{0,2}^y-\Omega_1^2\hat{\sigma}_{0,2}^x\bigg].
\end{equation} 

{In principle, one may introduce here a complex field correction to $\Omega_x$, as in Ref.~\cite{Li2024Experimental}; however, this will generate a large out-of-phase component and complicates the gate calibration due to large dynamical phases.}
Instead, we apply an additional virtual correction proportional to $\hat{\sigma}_{0,2}^y$, via $\hat{S}_3 = i\mathcal{E}^x_{02}2\Omega_1\dot\Omega_1 \hat{\sigma}_{0,2}^x/\Delta_2^2 $, 
that enables the resulting error term to be real~\cite{PhysRevA.88.062318, Wang2025Suppressinga}, leading to 
\begin{equation}
    \hat{\mathcal{H}}_3(t)  \approx\hat{\mathcal{H}}_1(t)
    - \mathcal{E}^x_{02} \bigg[2\frac{\Omega_1\ddot\Omega_1+{\dot\Omega_1}^2}{\Delta_2^2}+\Omega_1^2\bigg]\hat{\sigma}_{0,2}^x.
\end{equation} 
 
Collecting all terms involving \(\hat{\sigma}_{0,2}^x\) and setting them to zero 
yields the first recursive correction (R1D) pulse: 
\begin{equation}
    \Omega_x(t)=\sqrt{\Omega_1^2+\frac{2}{\Delta_2^2}(\dot{\Omega}_1^2+\ddot{\Omega}_1\Omega_1)}.
    \label{eq:1st_rec}
\end{equation}
\begin{figure}
    \centering
    \includegraphics[width=1\linewidth]{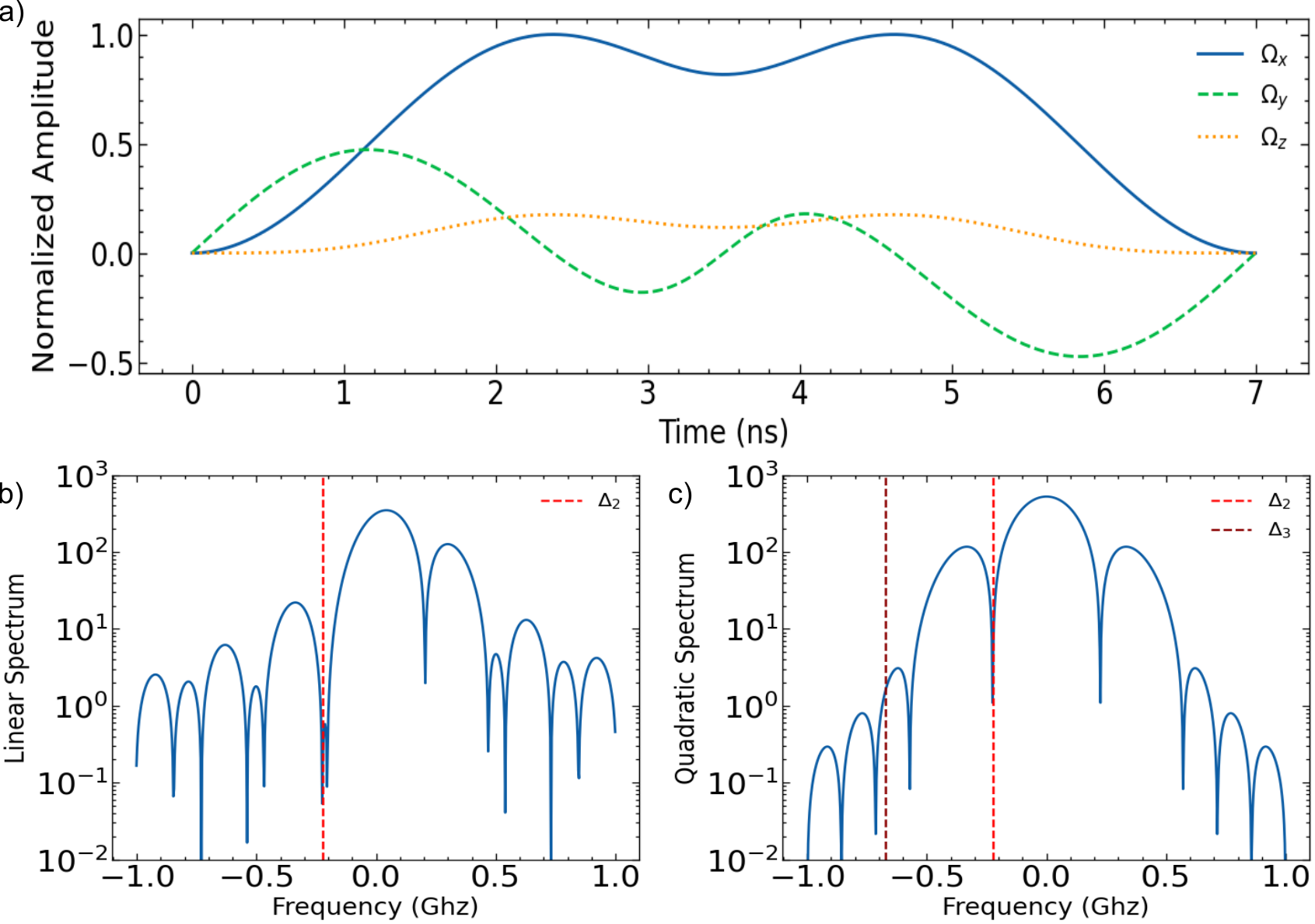}
    \caption{{\textbf{R1D pulse shape and Fourier spectrum of $\Omega$ and $\Omega_x^2$} a) Pulse shape of a R1D pulse starting with $\Omega_x= \Omega_0 \sin^3{(2\pi t/T)}$ for a gate time of $T= 7ns$. b) Fourier spectrum of a R1D pulse, $\Omega = \Omega_x + i \Omega_y$. The hole in the frequency spectrum associated with the $\ket{1}\rightleftarrows\ket{2}$ transition at $\Delta_2$ is preserved. c) Fourier spectrum of $\Omega_x^2$. As can be seen, we have introduced a hole at $\Delta_2$, suppressing the $\ket{0}\rightleftarrows\ket{2}$ transition. However the $\ket{1}\rightleftarrows\ket{3}$ transition has not yet been addressed and the increase in the power of the sin, from $sin^2$ to $sin^3$ has increased the power at this transition and therefore its importance.}  }
    \label{fig:r1d_pulseFT}
\end{figure}
We require that the R1D pulse satisfies the boundary conditions given in Eq.~(\ref{Exp2}). By direct inspection, $\Omega_{x}(t)$ vanishes at $t=0$ and $t=T$, provided that $\Omega_1$ obeys the same boundary conditions. However, $\dot{\Omega}_{x}(t)$ also needs to vanish at $t=0$ and $t=T$, which imposes a new constraint on $\Omega_{1}(t)$. Evaluating 
$        \lim_{t\to 0} \dot{\Omega}_x = 
        \lim_{t\to 0} \frac{\Omega_1+\frac{1}{\Delta_2^2}(3\ddot{\Omega}_1+\dddot{\Omega}_1\frac{\Omega_1}{\dot{\Omega}_1})}{\sqrt{(\frac{\Omega_1}{\dot{\Omega}_1})^2+\frac{2}{\Delta_2^2}(1+\ddot{\Omega}_1\frac{\Omega_1}{\dot{\Omega}_1})}}
$
and using \(\lim_{t\to 0}\Omega_1(t)/\dot{\Omega}_1(t) = 0\), the expression reduces to $\lim_{t\to 0} \dot{\Omega}_x = 3\ddot{\Omega}_1/\sqrt{2\Delta_2^2}$.
Thus satisfying the boundary conditions requires $\ddot{\Omega}_1(0)=\ddot{\Omega}_1(T)=0$. The Hann pulse does not meet this additional condition, so instead we use $\Omega_{1}(t)=\Omega_{0}\sin^3(\pi t/T)$. Relying on a higher power in the sinusoidal envelope does however sharpen the pulse in the time domain, which in turn broadens its frequency spectrum and potentially increases spectral content in unwanted transitions if no correction terms are added{, as can be seen in Fig. \ref{fig:r1d_pulseFT} c)}.

Using R1D, we implement a $\pi$-rotation along the $x$ axis. { It's infidelity as a function of gate time $T$ is shown in Fig.~\ref{fig:ana_fids}}. As can be seen, this technique outperforms standard DRAG for all times, but especially for extremely fast gates, achieving a gate fidelity above 99.99\% for gate times longer than $\approx$ 10.9~ns, a 14\% improvement in mimimum gate time.

{ In Fig. \ref{fig:r1d_pulseFT} we can see the pulse shape for a R1D pulse as well as its Fourier spectrum and the Fourier spectrum of the square of its real component. We have maintained the suppression of $\ket{1}\rightleftarrows\ket{2}$ and suppressed the $\ket{0}\rightleftarrows\ket{2}$ transition, as seen by the hole in the frequency spectrum of $\Omega_x^2$ at $\Delta_2$. However, $\ket{1}\rightleftarrows\ket{3}$ not only remains unaddressed, but slightly worsened, and needs to be suppressed. }

\subsection{Virtual leakage $\ket{1}\rightleftarrows\ket{3}$ and R2D pulse}
\label{R2D}

The remaining bottleneck arises from leakage to $\ket{3}$, a two-photon process that has not been previously analyzed.
After inserting the first recursion, the Hamiltonian restricted to the computational subspace and its transitions reads
\begin{equation}
     \hat{\mathcal{H}}_3(t) \approx\frac{1}{2} \Omega_x  \sigma_{0,1}^x+ \Omega_x^2 \mathcal{E}^x_{13}\hat{\sigma}_{1,3}^x.
\end{equation}

To suppress this remaining transition, we introduce a second recursive DRAG (R2D), targeting the $\ket{1}\rightleftarrows\ket{3}$ transition.

Because \(\Omega_x\) has already been defined as a functional of \(\Omega_1\), the next recursion must update \(\Omega_1\) itself.
Directly doing so is rather challenging due to its derivative terms introduced by Eq.~\eqref{eq:1st_rec}. We therefore perform a sequence of virtual transformations to eliminate these derivative contributions, which are frame transformations that do not modify the pulse shape. We introduce Schrieffer-Wolff generators with yet a new trial waveform {$\Omega_V$} so that
\begin{equation}
   \hat{S}_4(t) = -i\frac{\Omega_V^2 \mathcal{E}^x_{13}}{\Delta_3}\hat{\sigma}_{1,3}^y,\quad  \hat{S}_5(t) = -i\frac{2\Omega_V\dot \Omega_V \mathcal{E}^x_{13}}{\Delta_3^2}\hat{\sigma}_{1,3}^x, 
\end{equation}

which together yield:
 
\begin{equation}
    \hat{\mathcal{H}}_4(t) \approx \hat{\mathcal{H}}_3(t)
    - \mathcal{E}^x_{13} \bigg[2\frac{\dot{\Omega}_V^2+\ddot{\Omega}_V\Omega_V}{{\Delta_3}^2}+\Omega_V^2\bigg]\hat{\sigma}_{1,3}^x.
\end{equation}

\begin{figure}[t!]
    \centering
    \includegraphics[width=1\linewidth]{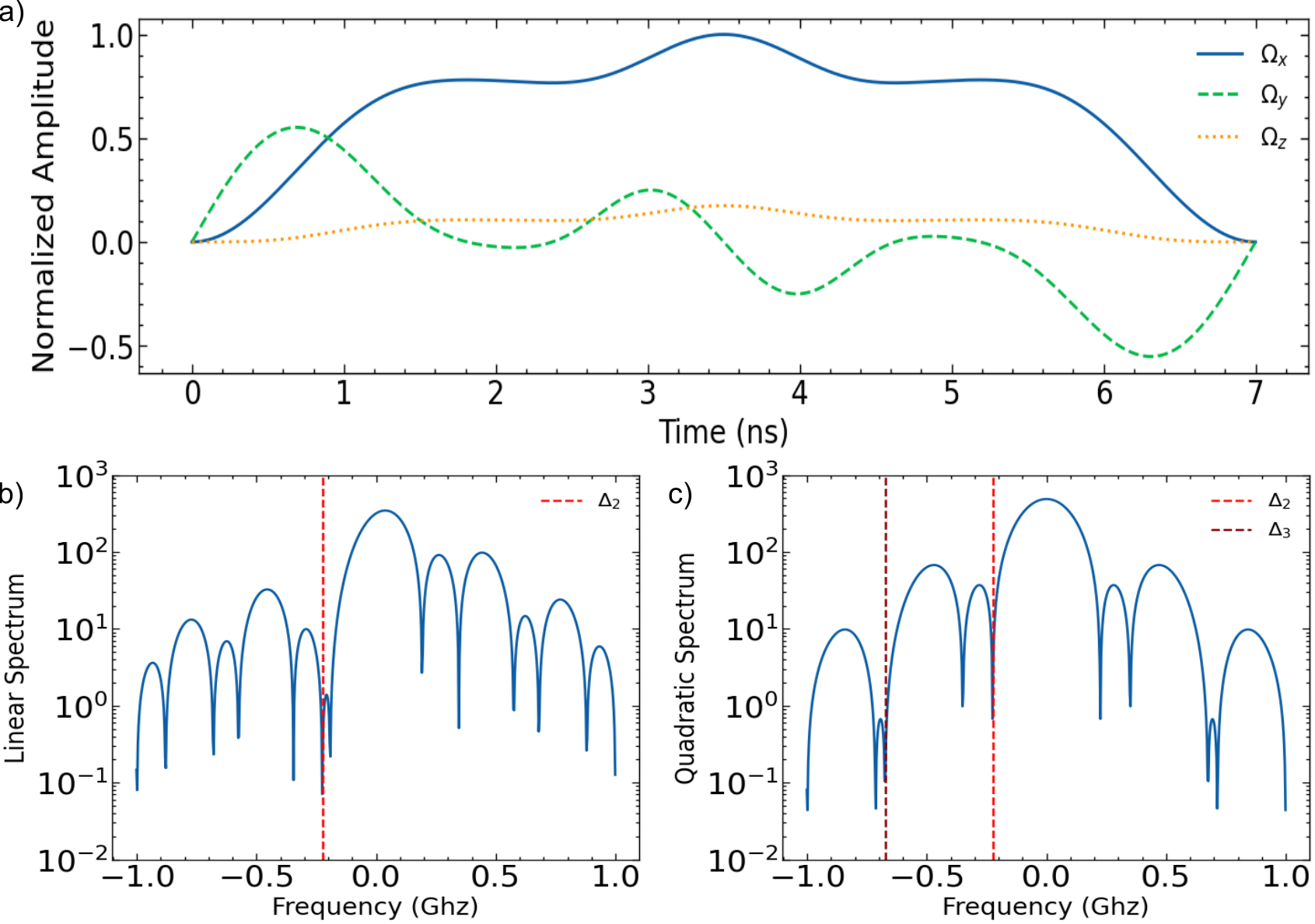}
    \caption{{\textbf{R2D pulse shape and Fourier spectrum of $\Omega$ and $\Omega_x^2$} a) Pulse shape of a R2D pulse starting with $\Omega_x= \Omega_I$ for a gate time of $T= 7ns$. b) Fourier spectrum of a R2D pulse, $\Omega = \Omega_x + i \Omega_y$. The whole in the frequency spectrum associated with the $\ket{1}\rightleftarrows\ket{2}$ transition at $\Delta_2$ is preserved. c) Fourier spectrum of $\Omega_x^2$. As can be seen, we now have a hole at $\Delta_2$ and $\Delta_3$ suppressing both $\ket{0}\rightleftarrows\ket{2}$and $\ket{1}\rightleftarrows\ket{3}$.}  }
    \label{fig:r2d_pulseFT}
\end{figure}
The role of these transformations is to eliminate derivatives of \(\Omega_1\) in the $\ket{1}\rightleftarrows\ket{3}$ transition,
 which leads  to the constraint

\begin{equation}
\frac{2}{\Delta_2^2}(\dot{\Omega}_1^2+\ddot{\Omega}_1\Omega_1)-\frac{2}{\Delta_3^2}(\dot{\Omega}_V^2+\ddot{\Omega}_V\Omega_V)+\Omega_V^2-\Omega_1^2=0.
\end{equation}
Choosing $\Omega_V = \Delta_3\Omega_1/\Delta_2$ reduces the remaining coupling in \(\mathcal{H}_4(t)\) to a single  \(\Omega_1^2\) term.
 
\begin{equation}
\hat{\mathcal{H}}_4(t) \approx\frac{\Omega_x}{2}  \sigma_{0,1}^x+ \Omega_1^2\mathcal{E}^x_{13}\bigg(\frac{\Delta_3^2}{\Delta_2^2}-1\bigg)\hat{\sigma}_{1,3}^x.
\label{eq:lastH}
\end{equation}
We now eliminate this term using a second recursive transformation using {a new} trial waveform, {$\Omega_2$}, given by the generators:
\begin{subequations}
\begin{eqnarray}
\label{Exp5}
\hat{S}_6(t)&=&-i\frac{\Omega_2^2}{\Delta_3}\mathcal{E}^x_{13}\bigg(\frac{\Delta_3^2}{\Delta_2^2}-1\bigg)\hat{\sigma}_{1,3}^y, \\
\label{Exp6}
\hat{S}_7(t) &=& i\frac{2\Omega_2\dot\Omega_2}{\Delta_3}\mathcal{E}^x_{13}\bigg(\frac{\Delta_3^2}{\Delta_2^2}-1\bigg)\hat{\sigma}_{1,3}^x,
\end{eqnarray}
\end{subequations}

yielding the second recursive (R2D) DRAG relation:
\begin{equation}
\label{recursion}
    \Omega_1(t) = \sqrt{\Omega_2^2+\frac{2}{\Delta_3^2}\big({\dot\Omega_2}^2+\ddot{\Omega}_2\Omega_2\big)}
\end{equation}

As before, we must determine the boundary conditions for the new trial pulse $\Omega_2$. We obtain that $\lim_{t\to 0} \ddot{\Omega}_1 = 3\dddot{\Omega}_2/\sqrt{2\Delta_3^2}$ which requires $\dddot{\Omega}_1(0)=\dddot{\Omega}_1(T)=0$.  A possible solution is $\Omega_{2}(t)=\Omega_{0}\sin^4(\pi t/T)$.

\subsection{Starting pulse shape}
Because of the sharpness of the new pulse envelope, we expect its frequency spectrum to be  broader and increasingly address other leakage transitions. Likewise, the higher the power in the pulse envelope, the larger the amplitude and consequently the higher order terms revealed by the DRAG frame may result in no improvement on the gate fidelity. 
However, this is not a limiting factor of the framework, instead it forces us to make an adequate selection of the initial pulse envelope that satisfies all the required boundary conditions and at the same time limits the maximal drive amplitude as much as possible. To this end, we  use the pulse proposed in Ref.~\cite{li2024universal},
\begin{eqnarray}
\label{eq:bl_pulse}
    \Omega_2(t)=\Omega_I(t) = \Omega_{{0}}\bigg[ \frac{1}{16}\cos{\frac{6\pi}{T}t}-\frac{9}{16}\cos{\frac{2\pi}{T}t}+\frac{1}{2}\bigg].
\end{eqnarray}
It is straightforward to demonstrate that $\Omega_I(t)$ satisfies all the boundary conditions required by both R1D and R2D. Moreover, this pulse has the advantage that the maximal amplitude is smaller than the  extension of the Hann pulse as $\sin^4(\pi t/T)$.

{Fig.~\ref{fig:fig1_inks} (d) shows the error budget for the different pulses, and that for ultra-fast gates the need becomes apparent to remove the $\ket{1}\rightleftarrows\ket{3}$ transition. If we do not remove it, this becomes the limiting factor. By using the second square root we can obtain a decrease by more than an order of magnitude in the population of $|3\rangle$.}

{ In Fig. \ref{fig:r2d_pulseFT} we can see the pulse shape for a R2D pulse as well as its Fourier spectrum and the Fourier spectrum of the square of its real component. We have maintained the suppression of $\ket{1}\rightleftarrows\ket{2}$ and the suppression of $\ket{0}\rightleftarrows\ket{2}$, as seen by their respective holes in the frequency spectrum of $\Omega$ and $\Omega_x^2$. Additionally, we introduced a hole at $\Delta_3$ in the frequency spectrum of $\Omega_x^2$ suppressing $\ket{1}\rightleftarrows\ket{3}$. }

\subsection{Superlinear corrections within the qubit manifold}
\label{SupraLinear}
We have demonstrated that it is possible to systematically reduce the error for single-qubit gates by including different recursions of the DRAG framework leading to significant error reduction. However, it is possible to reduce  the error even more, achieving fidelities around $99.999\%$,  by including corrections that involve higher order contributions to the X and Y component of the pulses. These corrections are obtained by expanding $\hat{\mathcal{H}}_4(t)$ Eq.~(\ref{eq:lastH})  up to third order in $\Omega_x/ \Delta_2$, whose form is shown explicitly in Appendix~\ref{supralinear_corrections_annex},  Eq.~\ref{eq:finalH}, or in simplified notation:

\begin{eqnarray}
    \hat{\mathcal{H}}_5(t) &\approx& \bigg[\frac{\Omega_x}{2}+\mathcal{E}^x_{01}\Omega_x^3\bigg]\hat{\sigma}_{0,1}^x\\&&+\Omega_x(\mathcal{E}^{y1}_{12}\Omega_1\dot\Omega_1+\mathcal{E}^{y2}_{12}\Omega_2\dot\Omega_2)\hat{\sigma}_{1,2}^y \nonumber \\
    &&+ \Omega_x(\mathcal{E}^{x0}_{12}\Omega_x^2+\mathcal{E}^{x1}_{12}\Omega_1^2+\mathcal{E}^{x2}_{12}\Omega_2^2)\hat{\sigma}_{1,2}^x \nonumber \\
    &&+ \Omega_x(\mathcal{E}^{y1}_{03}\Omega_1\dot{\Omega}_1+ \mathcal{E}^{y2}_{03}\Omega_2\dot{\Omega}_2)\sigma_{0,3}^y \nonumber \\
    &&
    + \Omega_x(\mathcal{E}^{x0}_{03}\Omega_x^2+\mathcal{E}^{x1}_{03}\Omega_1^2+\mathcal{E}^{x2}_{03}\Omega_2^2)\sigma_{0,3}^x + O(\Omega^4)
\nonumber
\end{eqnarray}

To avoid over/under rotation on the computational subspace, we need to modify the pulse envelope to cancel the higher order correction $\mathcal{E}^x_{01}\Omega_x^3$ by updating the pulse with
\begin{equation}
    \Omega_x^{'}(t) = \Omega_x(t)-\frac{(4-\lambda_2^2)\Omega_x^3(t)}{8\Delta_2^2}.
    \label{eq:Cube}
\end{equation}
The  remaining higher order terms in $\ket{1}\rightleftarrows\ket{2}$ can be removed by changing the Y component of the pulse, which needs to be updated according to the effective Hamiltonian obtained through two transformations. The first is a modification of initial DRAG transformation \ref{eq:frame_DRAG}:

\begin{equation}
    \hat{S}_{8}(t) = -i\sum_{j=1}^{m} \frac{\lambda_j}{2}  \bigg[\frac{\Omega_x}{\Delta_2}+\frac{A_{0,1}(t)}{\Delta_2^3}\bigg] \hat{\sigma}^y_{j-1, j}
    \label{eq:Agenerator}
\end{equation}
where $A_{0,1}(t)$ is an auxiliary variable defined in  Appendix~\ref{supralinear_corrections_annex} Eq. \ref{eq:Adef}, and must be applied as the first step in the analysis. This frame transformation will induce an extra term $-\frac{\lambda_2}{2\Delta_2^2} A_{0,1}(t) \hat{\sigma}_{1,2}^x$ in Eq. \ref{eq:finalH} that can be used to remove the extra terms in $\hat{\sigma}_{1,2}^x$.
The second frame transformation changes the terms in $\hat{\sigma}_{1,2}^y$ to terms in $\hat{\sigma}_{1,2}^x$ so that all superlinear terms may be corrected and it is defined as:

\begin{equation}
    \hat{S}_{9}(t)=-\frac{i}{\Delta_2}\Omega_x(\mathcal{E}^{y1}_{12}\Omega_1\dot\Omega_1+\mathcal{E}^{y2}_{12}\Omega_2\dot\Omega_2)\hat{\sigma}_{1,2}^x
    \label{eq:Aswitch}
\end{equation}

To remove both the Y component introduced in the pulse and the Stark-shift we perform the following pulse updates:
\begin{equation}
    \Omega_y^{'}(t)=-\frac{\dot{\Omega^{'}}_x(t)}{\Delta_2}-\frac{\dot{A}_{0,1}(t)}{\Delta_2^3},\quad \delta^{'}=\delta-\frac{A_{0,1}(t)\Omega_x}{\Delta_2^3}.
    \label{eq:Ycube}
\end{equation}

These corrections in Eq.~(\ref{eq:Cube} and \ref{eq:Ycube}) are not exclusive to R2D and can be performed to every DRAG framework. In practice, they can also be corrected by optimizing the amplitude of pulses in experiments. Therefore, the corrections were included in all frameworks in Fig.~\ref{fig:ana_fids} to demonstrate a fair comparison of all methods under such higher order optimization. Note that for linear DRAG these expressions simplify substantially and $A_{0,1}(t)= (4-\lambda_2^2)\Omega_x^3(t)/(8\Delta_2^2)$. This can be implemented by using $\Omega_x^{'}$ and simply taking $\Omega_y^{'}(t)=-\dot{\Omega}_x(t)/\Delta_2$ and therefore no Y correction is necessary.
 
 Using the corrected R2D pulse, we obtain errors an order of magnitude smaller than standard DRAG above 6ns and R1D above 12ns.  
 As can be seen in Fig. \ref{fig:ana_fids} (c), a fidelity of 99.999\% is obtained for pulses from 12~ns. 

 After the correction, we obtain a remaining leakage term involving the three-photon process in $\ket{0}\rightleftarrows\ket{3}$:

\begin{eqnarray}
    &&\hat{\mathcal{H}}_6(t) \approx \bigg[\frac{\Omega_x'}{2}\bigg]\hat{\sigma}_{0,1}^x+ \Omega_x(\mathcal{E}^{y1}_{03}\Omega_1\dot{\Omega}_1+ \mathcal{E}^{y2}_{03}\Omega_2\dot{\Omega}_2)\sigma_{0,3}^y \nonumber \\
    &+& \Omega_x(\mathcal{E}^{x0}_{03}\Omega_x^2+\mathcal{E}^{x1}_{03}\Omega_1^2+\mathcal{E}^{x2}_{03}\Omega_2^2)\sigma_{0,3}^x + O(\Omega^4)
\end{eqnarray}

These terms are difficult to remove and cannot be easily simplified by virtual transitions due to the fact that all terms are proportional to $\Omega_x$ and $\Omega_x$ is dependent on $\dot{\Omega}_1$, $\ddot{\Omega}_1$
$\dot{\Omega}_2$, $\ddot{\Omega}_2$.\\

\subsection{Minimum Gate time}

{In  Eq.~(\ref{eq:1st_rec}) we have expressed the real component of our pulse, $\Omega_x$, in terms of a square root. The terms in this square root have to be positive during the gate duration, since the imaginary component of the pulse, $\Omega_y$, has already been defined. Due to the second derivative, we introduce negative components that will establish a minimum gate time for a full rotation of the R1D pulse considered, $T_{{\rm{min}}} = \sqrt{6}\pi/\Delta_2$}.

We can generalize this expression to several orders of recursion by assuming that the initial pulse has as shape $\Omega_{1}(t)=\Omega_{0}\sin^n(\pi t/T)$ with the condition $\Omega_1^2(T_{{\rm{min}}}) + 2[\dot{\Omega}^2_1(T_{{\rm{min}}}) +\ddot{\Omega}_1(T_{{\rm{min}}})\Omega_1(T_{{\rm{min}}})]/\Delta_2^2 =0$. By replacing the argument of $\Omega_{x}(t)$ from Eq.~(\ref{eq:1st_rec}), we get
\begin{equation}
\left[1 - \frac{2n\pi^{2}}{\Delta_{2}^{2} T_{\min}^{2}}\right]
\sin^{2}\!\left[\frac{\pi t}{T_{\min}}\right]
+ \frac{(4n^2-2n)\pi^{2}}{\Delta_2^2T_{\min}^{2}}
\cos^{2}\!\left[\frac{\pi t}{T_{\min}}\right] = 0.
\label{eq:min_gate_t}
\end{equation}
We need to guarantee that this expression is positive for all times. However, we know that the minimum of Eq.~(\ref{eq:min_gate_t}) is achieved when $t=T_{{\rm{min}}}/2$ {(so that $\cos^{2}\!\left[\frac{\pi t}{T_{\min}}\right]=0$ and we no longer have the terms proportional to $\cos^2$, which are always positive)}, reducing the expression to
\begin{equation}
    T_{{\rm{min}}}=\sqrt{2n}\frac{\pi}{\Delta_2}.
\end{equation}
For the R2D pulse, we require that both square roots remain positive. Following the same procedure as above we obtain the following:

\begin{eqnarray}
    T_{{\rm{min}}}= \pi\sqrt{\frac{n}{\Delta_3^2}+\frac{n}{\Delta_2^2} +\sqrt{\frac{n^2}{\Delta_3^4}+\frac{n^2}{\Delta_2^4}+\frac{4n-10n^2}{\Delta_2^2\Delta_3^2}}}
\end{eqnarray}
which reduces to the previous equation under the assumption $\Delta_3 \gg \Delta_2$. This leads us to the interesting and unexpected finding that, for the same $\Delta_2$, the minimum gate time $T_{min}$ for R2D increases with increasing $\Delta_3$, that is, the bigger the separation between $|2\rangle$ and $|3\rangle$ for the same anharmonicity, the slower the gate {(for the same n). This is a consequence of the first recursion reducing the amplitude of $\Omega_1$, when compared to $\Omega_2$, which allows for greater values of $\ddot{\Omega}_2$.  However, if $\Delta_3 \gg \Delta_2$, $\Omega_1 \approx \Omega_2$ and the first recursion has no effect on pulse shape.}

The minimum gate time for R1D and R2D using the pulse in Ref.~\cite{li2024universal} is obtained by ensuring both square roots are positive. This can be solved numerically and, for an anharmonicity of $\Delta_{2}/2\pi=-0.225$ GHz gives us a gate time of $\approx$ 4.45~ns.

\section{Prefactor values from analytical time-ordering and numerical optimization}\label{Numerics}

It is now common practice to optimize the coefficient for linear DRAG in single qubit gates~\cite{PhysRevLett.116.020501, marxer2025999fidelitysinglequbitgates}.
Yet the underlying physics governing these optimized corrections is seldom clarified.
This gap arises both from discrepancies between the idealized model and the actual device, such as parameter drift~\cite{Burnettdriftt1t2,graaffcali,wudrift}, distortion on the pulses~\cite{PhysRevLett.110.040502,PhysRevApplied.21.064060} andcrosstalk~\cite{Wang2025Suppressinga}, and from the presence of additional destructive interference pathways among different error channels.
While treating the calibration as a black-box optimization has enabled state-of-the-art fidelities, further progress entails a deeper understanding of the dynamics.

In this section, we analyze the role of each free parameter and develop a systematic framework for Hamiltonian engineering. Our analysis reveals how these parameters interact, identifies which error channels they influence, and provides analytical expressions for their near-optimal values.

\subsection{Parameterization of DRAG solution families} \label{drag_num}

For linear DRAG, a set of tunable parameters is given by the triad $\{\beta, \alpha, \delta_{c}\}$, first introduced in \cite{PhysRevA.83.012308}, that appears in the control as

\begin{eqnarray}
\Omega \rightarrow \beta (\Omega_x+\alpha \Omega_y),\quad \delta \rightarrow \delta_c.
\label{eq:num_drag}
\end{eqnarray}
where $\beta$ parameterises changes in the total amplitude and $\alpha$ the ratio for the Y component. The parameter $\delta_{c}$ is a constant detuning correction due to frequency mismatch between the carrier frequency of the drive and the real transition frequency of the system, introduced by the AC-Stark shift. This set of parameters follows the one used in Ref.~\cite{PhysRevLett.116.020501}.

Fig.~\ref{fig:fig1_inks} c) compares the gate infidelity $\mathcal{E}(\hat{\mathcal{U}}_Q)$ as a function of the gate time for different pulses. We observe that the optimized version of linear DRAG { (represented by the acronym DRAG N, the green dashed line)} always outperforms the analytical one for gate times longer than 6~ns {(the blue line)}. We find that around 9.1~ns we have a destructive interference between the remaining uncorrected leakage channels, such that we can eliminate all leakage remaining. This way,  between $\text{T}=8.93$~ns and $\text{T}=9.33$~ns  we achieve the threshold of error bellow $10^{-5}$, assuming sufficiently low decoherence. Moreover, we observe that this numerical optimization exhibits quasi-periodic peaks that are monotonically decreasing in the gate fidelity, though these are outside the range of the region of interest that is plotted, indicating an effective recurrence time. Note however that picking the precise time duration for a low-infidelity gate is known \cite{Warrenmagnus} to be problematic for a number of reasons. Aside from pulse timing errors, the calibration process becomes much more cumbersome as in addition to the calibration of the pulse parameters, the gate duration must also be calibrated. In a multi-qubit processor, these optimal durations will moreover vary from qubit to qubit, which prohibits a common clock frequency for the processor, rendering quantum circuits highly cumbersome. Therefore, one cannot in general count on such an approach for ultra-low error.

{In the next sections, we now aim to }further understand the underlying physics behind these parameters{, and why such an increase in performance is gained.} 
{Therefore, we } modify Eq. (\ref{eq:frame_DRAG}) to consider a continuous set of adiabatic transformation given by \cite{PhysRevA.83.012308}
\begin{equation}
    \hat{S}_D(t) = -i \lambda_1\frac{\alpha\Omega_x}{2\Delta_2} \hat{\sigma}^y_{0,1} -i\sum_{j=2}^{m}  \lambda_j \frac{\Omega_x}{2\Delta_2} \hat{\sigma}^y_{j-1, j}.
    \label{eq:framealpha}
\end{equation}
We use this frame to ensure no terms proportional to $\hat{\sigma}_{0,1}^y$ remain in the computational subspace whilst still eliminating the leakage term $\hat{\sigma}_{1,2}^x$.
\begin{figure*}[!t]
    \centering
    \includegraphics[width=0.9\linewidth]{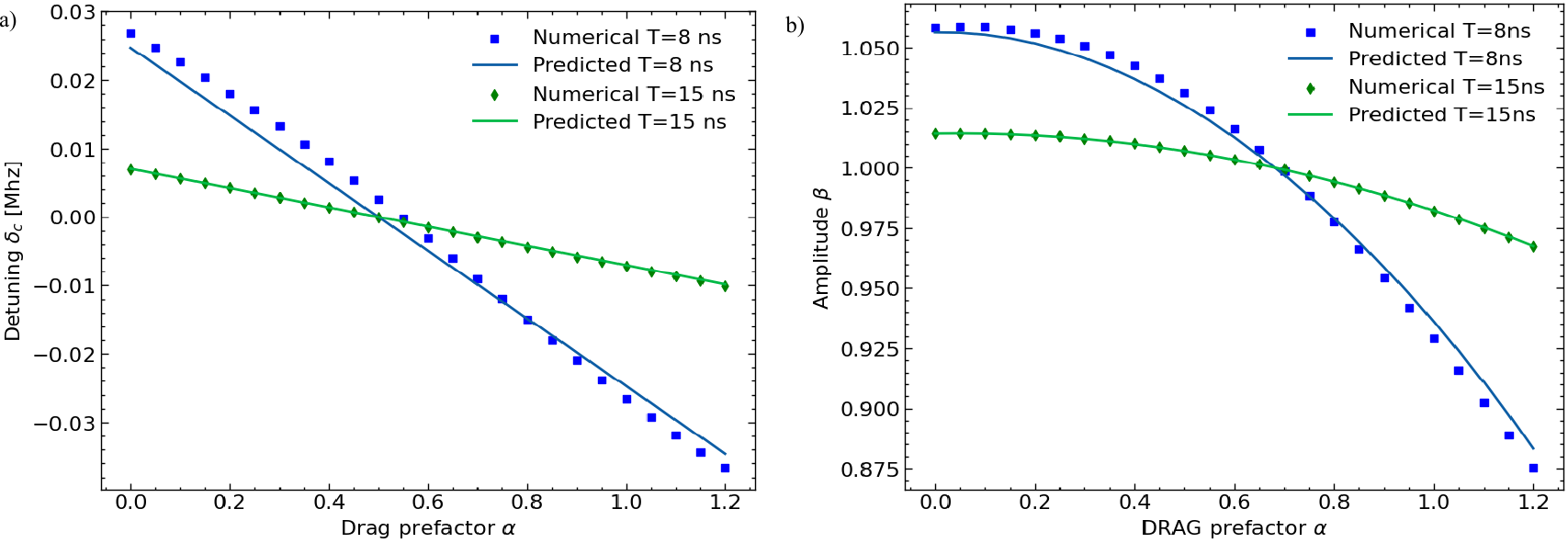}
    \caption{ \textbf{Amplitude and Constant Detuning as a function of $\alpha$} a) Comparison between the optimized values of constant detuning $\delta_c$ for the DRAG pulse and the analytical approximation, for different values of $\alpha$. The analytical expression can accurately predict the values of constant detuning. The longer the gate time, the better the prediction. b) Comparison between the optimized values of $\beta$ for  the DRAG pulse and the analytical approximation, for different values of $\alpha$. Once again, the analytical expression can accurately predict the values of constant detuning. }
    \label{fig:drag_betas}
\end{figure*}
\subsection{Optimal detuning}

We update the triad $\{\beta, \alpha, \delta_{c}\}$ such that  both leakage and phase errors cancel at the same time. To do so, we transform the starting Hamiltonian, Eq. (\ref{Exp:RWA}), and expand to third order in $O(\Omega_x/\Delta)$, using the new frame generator, Eq. (\ref{eq:framealpha}). Additionally,  we move to the \textit{toggling frame}~\cite{Haeberlen1976,Brinkmann2016} given by the unitary which will represent ideal dynamics:
\begin{equation}
    \hat{V}_{{\rm{tog}}}(\theta)=\exp(-i\theta\sigma^x_{0,1}+\phi_1|2\rangle\langle 2| + \phi_3 |3\rangle\langle3|)
\end{equation}
where the rotation angle is given by $\theta(t)=\int_{0}^{t}\Omega_{x}(s)ds/2$ and $\phi_1$ and $\phi_2$ allow for free precession beyond the qubit subspace. Within the qubit subspace, this gives
\begin{eqnarray}
   \hat{\mathcal H}^{(qub)}_{\rm{tog}}  &=& \left(\left[\delta_c+\mathcal{E}_{11}\Omega_x^2\right]\cos{\theta}+\mathcal{E}_{11}^{(\alpha)}\Omega_x\dot{\Omega}_x\sin{\theta}\right)\ket{1}\bra{1}\nonumber \\
    &&+\bigg(\Omega_x-\dot\theta+i\left[\delta_c+\mathcal{E}_{11}\Omega_x^2\right]\sin{\theta} - \mathcal{E}^{x0}_{01}\Omega_x^3\nonumber \\
    &&+\mathcal{E}^{x1}_{01}\Omega_x\delta_c+i\mathcal{E}_{11}^{(\alpha)}\Omega_x\dot{\Omega}_x\cos{\theta}\bigg)\frac{\sigma^x_{0,1}}{2},~~
    \label{eq:Htog}
\end{eqnarray}
where the $\mathcal{E}$ prefactors are given in Appendix \ref{Magnus_App}, as well as remaining transitions to higher levels.
It is important to note that $\hat{V}_{{\rm{tog}}}(T) \neq I$, and in fact $\hat{V}_{{\rm{tog}}}(T)$ generates the desired $\pi$ rotation. Therefore, the transformed toggling Hamiltonian will represent the error of our gate. If $H_{{\rm{tog}}} =0$ for all time, the $\pi$ rotation can be ideally implemented. For now we ignore the effect of $\beta$ noting that for $\beta=1$ we have $\Omega_x=\dot\theta$.

The goal of the toggling transformation is to separate the error part of the dynamics $H_{{\rm{tog}}}$, which is typically small. This ensures fast convergence of the Magnus expansion~\cite{Warrenmagnus,Blanes_2009,PRXQuantum.6.010328} which allows us to express the propagator as  $\hat{\mathcal{T}}\exp(-i\int_0^T \hat{\mathcal{H}}_{\rm{tog}}(t) dt)=\exp(-i\sum_{\ell}\hat{\mathcal{Z}}_{\ell}(t))$, where $\hat{\mathcal{Z}}_{\ell}(t)$ are the different orders of the expansion defined as 
\begin{subequations}
\begin{eqnarray}
\label{FirstOrderMag}
\hat{\mathcal{Z}}_{1}(t)&=&\int_{0}^{T}dt\hat{\mathcal{H}}_{\rm{tog}}(t),\\
\label{SecondOrderMag}
\hat{\mathcal{Z}}_{2}(t)&=&\int_{0}^{T}dt\int_{0}^{t}d\tau\big[\hat{\mathcal{H}}_{\rm{tog}}(t),\hat{\mathcal{H}}_{\rm{tog}}(\tau)\big],
\end{eqnarray}
\end{subequations}
Thus, we shall find the conditions for the parameters introduced in the toggling frame to make $\hat{\mathcal{Z}}_{1}(t)$ close to 0 as possible.

For $\pi$ rotation, the phase error on $\ket{1}\bra{1}$ turns out to be trivially zero due to symmetry. This can be seen because each of the terms in the first line of  Eq.~\ref{eq:Htog} is anti-symmetric around the half-time of the operation when $\theta$ undergoes the gate  period (see also Appendix~\ref{Magnus_App}).  However, in general, one would not be able to make both equations zero using constant detuning. A time-dependent detuning, like the one used in Section \ref{analytical section} would be required. This also explains why using the DRAG formula with constant detuning for partial rotation is often seen to have larger error than one would naively expect given the proportionate reduction in pulse amplitude.

The requirement that the $\sigma^x_{0,1}$  term also integrates to zero gives us the  tools to predict the value of the constant detuning and the amplitude of the pulse. Zeroing the imaginary part (c.f.~Eq.~\ref{eq:err01}), we get the equation for $\delta_c$:
\begin{equation}
    \delta_c = -\frac{\int_0^T\mathcal{E}_{11}\Omega_x^2\sin{\theta}}{\int_0^T\sin{\theta}}-\frac{\int_0^T\dot{\Omega}_x\mathcal{E}^{(\alpha)}_{11}\Omega_x\cos{\theta}}{\int_0^T\sin{\theta}}.
    \label{eq:cnt_det_comp}
\end{equation}
Note that this is not just an average of the accumulated time-dependent Stark shift, but is renormalized by the rotation angle $\theta$.
Assuming a pulse shape of the form $\Omega_x=2\pi\sin^2({\pi t/T})/T$ we get:
\begin{equation}
    \delta_{c}\approx0.712\frac{\lambda_2^2-4\alpha}{\Delta_2}\frac{\pi^2}{T^2}. 
    \label{eq:cnt_det}
\end{equation}

Fig.~\ref{fig:drag_betas} (a) compares the analytical expression using Eq.~\ref{eq:cnt_det} with numerically optimized results for different DRAG coefficient $\alpha$ and two different gate times $T$=8~ns and $T$=15~ns. The level of agreement is governed by the Magnus expansion: as the gate time increases, higher-order terms {(whose terms have more commutators of $\hat{\mathcal{H}}_{\rm{tog}}$ and terms proportional to higher powers of $\Omega_x/\Delta_2$)} become negligible, and the analytical and numerical results converge. The constant detuning expression also explains the physics observed in the experimental realization in Ref.~\cite{PhysRevLett.116.020501}, where a linear relationship between $\alpha$ and $\delta_{c}$ was reported with $\delta_{c}=0$ roughly at $\alpha=0.5$ (for $\lambda_2\approx\sqrt{2}$) \cite{PhysRevA.83.012308}. The experiment also observed scaling $\delta_{c} \propto 1/T^2$, which is verified by our analysis. Recent results in Ref.~\cite{GAO2025} further confirm the strong agreement between this analytical expression and calibrated constant-detuning values.

\subsection{{Optimal Pulse Amplitude}}

 Taking the real part of the integral (Eq.~\eqref{eq:err01}), we have the following term for the amplitude of the pulse:
\begin{equation}
    \int_0^T dt\left[\beta\Omega_x-\dot{\theta}-\beta\Omega_x(\mathcal{E}^{x1}_{01}\delta_c+\mathcal{E}^{x0}_{01}\beta^2\Omega_x^2)\right]
    \label{eq:beta_3}
\end{equation}
which is a cubic polynomial of $\beta$ and must be set to zero.
Despite Eq.~(\ref{eq:beta_3}) having an analytical solution, it is quite cumbersome to solve. Instead, assuming a mean-field approximation as $\beta=1+\eta$ by considering a small variation of the expected optimal amplitude, we can linearize Eq.~(\ref{eq:beta_3}) as $\beta^3\approx1+3\eta$, leading to
\begin{equation}
    \beta =1 +\frac{16\alpha\delta_c\pi\Delta_1+(4\alpha^2+\lambda_2^2-2\alpha\lambda_2^2)\frac{5\pi^3}{T^2}}{16\pi\Delta_1^2-16\alpha\delta_c\pi\Delta_1-3(4\alpha^2+\lambda_2^2-2\alpha\lambda_2^2)\frac{5\pi^3}{T^2}},
    \label{eq:beta_app}
\end{equation}
corresponding to the correction of the drive amplitude as a function of the DRAG coefficient $\alpha$ for fixed gate time $T$. It also shows how one can correct the X superlinear correction identified in section \ref{SupraLinear} alternatively, through prefactor calibration.

Fig.~\ref{fig:drag_betas} (b) compares the amplitude correction $\beta$ obtained in Eq.~(\ref{eq:beta_app}) with the one obtained through numerical optimization at two different gate times. Just as in the previous case, the agreement between the analytical and the numerical results is conditioned on the validity of the first-order Magnus expansion and improves with gate times, and is generally extremely good.

\subsection{Linear DRAG prefactor}\label{alpha_pred}

Finally, to understand and predict the value of $\alpha$ we must look into {the $|1\rangle\langle 2|$ term of $\hat{\mathcal{Z}}_{1}$, $\hat{\mathcal{Z}}_{1}[1,2]$ (and the term $|2\rangle\langle1|$,  $\hat{\mathcal{Z}}_{1}[2,1]$) and the $|0\rangle\langle 2|$ term of $\hat{\mathcal{Z}}_{1}$, $\hat{\mathcal{Z}}_{1}[0,2]$ (and the term $|2\rangle\langle0|$, $\hat{\mathcal{Z}}_{1}[2,0]$)} which represent the main leakage channel in our error Hamiltonian (and whose contributions came from a mixture of the $\ket{1}\rightleftarrows\ket{2}$ and $\ket{0}\rightleftarrows\ket{2}$ transitions in the lab frame). These elements can be found in Appendix~\ref{Magnus_App}, Eq. \ref{eq:Z12} and Eq. \ref{eq:Z02}.

In order to minimize the leakage we now want the $\alpha$ that minimizes  $l=\abs{\hat{\mathcal{Z}}_{1}[1,2]}^2+\abs{\hat{\mathcal{Z}}_{1}[0,2]}^2$. This can be found by taking the derivative $\partial l/\partial \alpha =0$. To solve this equation, we approximate $\phi_\ell=\Delta_{\ell}t$, and we once again perform a mean field approximation, this time over the alphas $\alpha=(1+\delta \alpha)$ such that $\alpha^n=(1+n\delta \alpha)$.

\begin{figure}[t]
    \centering
    \includegraphics[width=0.85\linewidth]{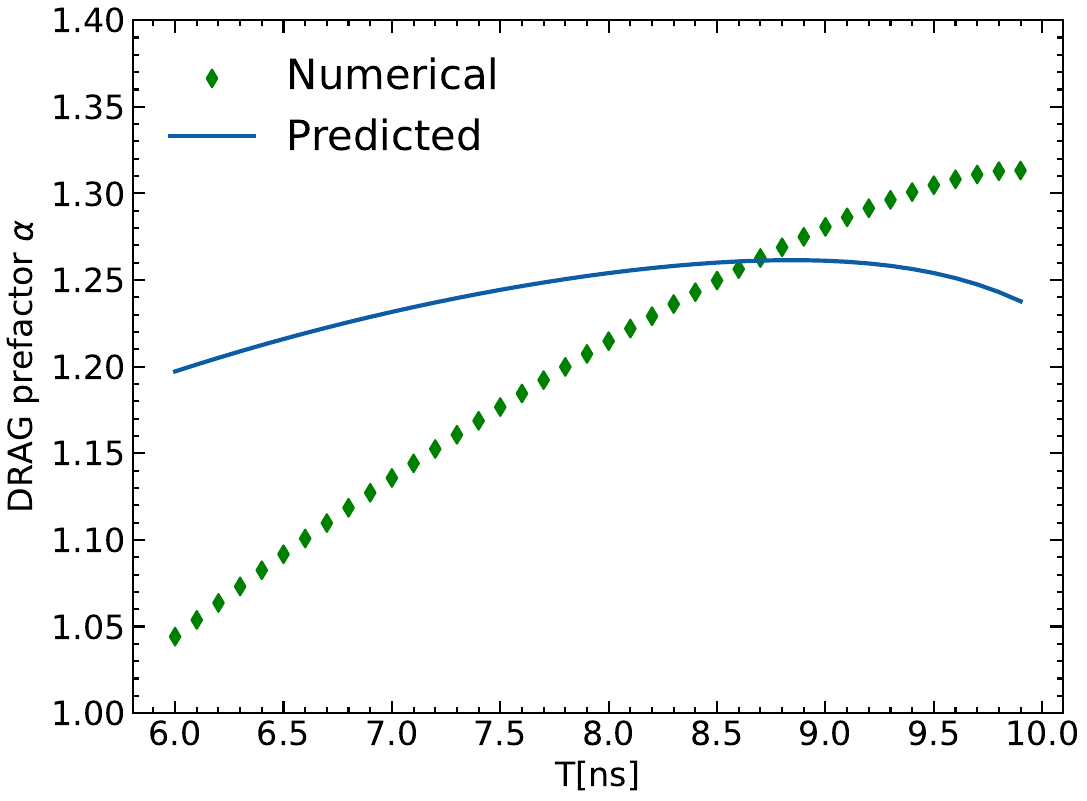}
    \caption{\textbf{Analytically vs Numerically optimized $\boldsymbol{\alpha}$ parameters.}  Comparison between the optimized values of $\alpha$ for linear DRAG pulse and the analytical expression of $\alpha$, as a function of gate time T. Although our expression captures the general behaviour of $\alpha$, we find large discrepancies, most likely due to higher orders. }
    \label{fig:alpha1}
\end{figure}
Solving this equation we obtain the following expression for $\alpha$:
\begin{equation}
    \alpha \approx 1-\frac{x_1+\Bar{x}_1}{x_2+\Bar{x}_2},
    \label{eq:approx_alphas}
\end{equation}
with $x_1$ and $x_2$ defined in Appendix~\ref{Magnus_App}.

\begin{figure}[!b]
    \centering
    \includegraphics[width=0.85\linewidth]{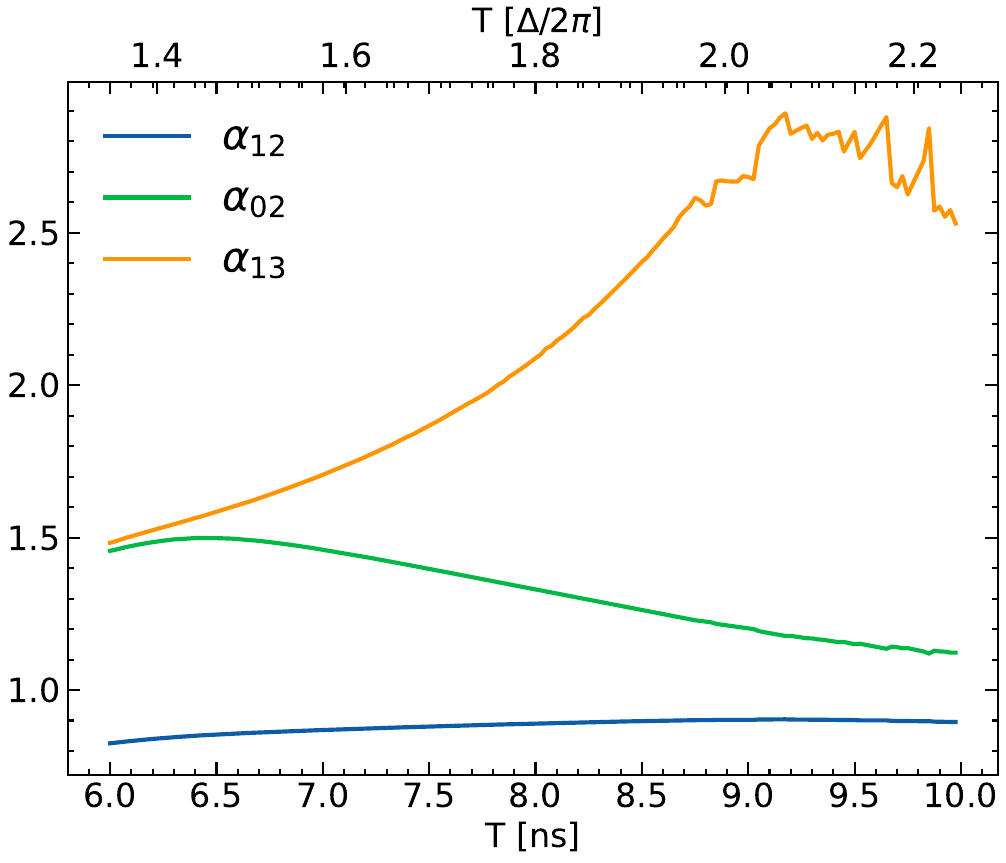}
    \caption{\textbf{ Numerically optimized $\alpha$ parameters.}     Optimal values of $\alpha_{12}$, $\alpha_{02}$ and $\alpha_{13}$ for the R2D pulse as a function of gate time T. These results generally match the ones obtained in experiment \cite{GAO2025}, providing validation of the theory.}
    \label{fig:all_alphas}
\end{figure}

Fig. \ref{fig:alpha1} compares the analytical expression for $\alpha$ with values obtained from full numerical optimization. In principle, $\alpha=1$ should completely compensate for the single-photon leakage channel $\ket{1}\rightleftarrows\ket{2}$. The analytical prediction lies around 1.2, reflecting the trade-off between this channel and the concurrent $\ket{0}\rightleftarrows\ket{2}$ transition and explicitly demonstrating why $\alpha\neq1$.
Numerical optimization reveals a steeper slope for the $\alpha$-dependence, ranging from 1.0 to 1.3, which arises from higher-order corrections not included in the analytical calculation. {The fact that the optimal $\alpha$ is not 1 was first experimentally demonstrated in \cite{PhysRevLett.116.020501} and it is here explained.}

With the prediction of all three parameters, we can now compare the performance between the predicted prefactors and those from full numerical optimization. 
{We plot the infidelity obtained for the DRAG pulse using our predictions of the 3 constants, $\alpha$, $\beta$ and $\delta_c$, in Fig. \ref{fig:fig1_inks} c), respectively termed DRAG P.} As seen, despite the discrepancy in the $\alpha$, the predicted prefactors provide large improvements to gate fidelity, with more than 1 order of magnitude. The largest deviation in fidelity is observed around 9~ns, where destructive interference is present among different leakage channels. In this regime, numerical optimization achieves further gains because it fully explores these interference pathways by optimizing the final unitary directly. {Nevertheless, the expressions here obtained can be used as a warm start for further calibration in an experimental setting}

\begin{figure*}[!t]
    \centering
    \includegraphics[width=0.90\linewidth]{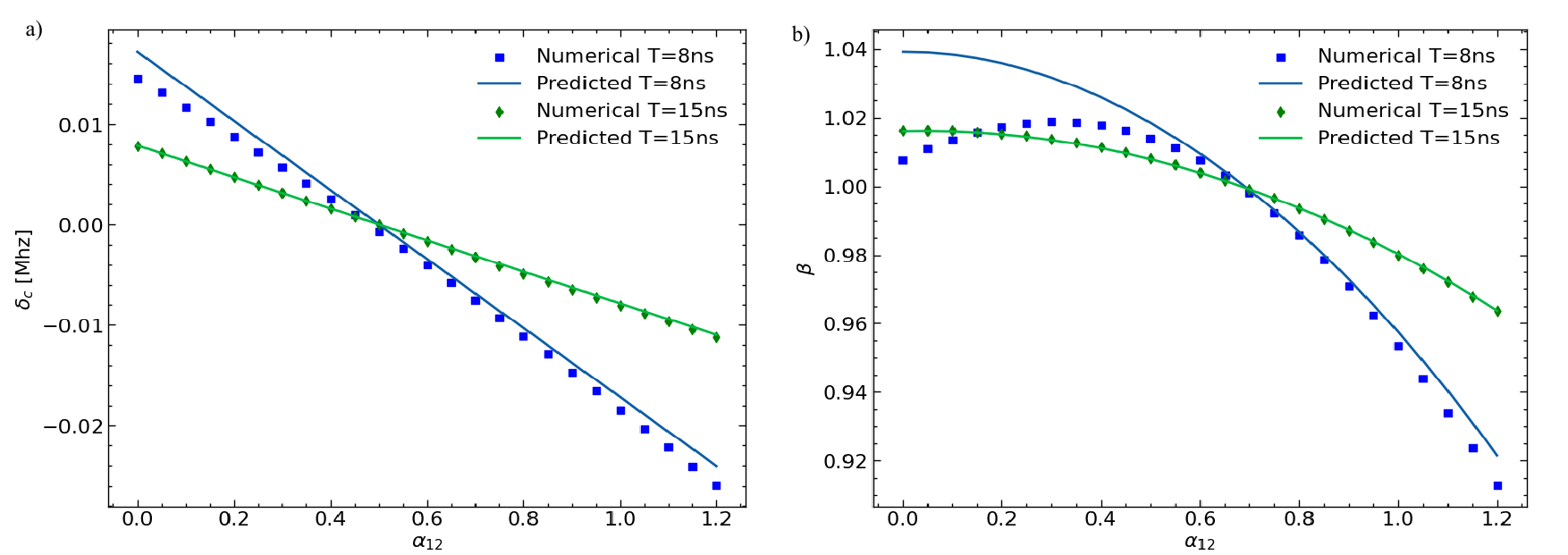}
    \caption{\textbf{Amplitude and Constant Detuning as a function of $\alpha_{12}$} a) Comparison between the optimized values of constant detuning for the R2D and the analytical approximation, for different values of $\alpha_{12}$. To make this example, we chose $\alpha_{02}=1.3$ and $\alpha_{13}=1.7$. The analytical expression can accurately predict the values of constant detuning. The longer the gate time, the better the prediction. b) Comparison between the optimized values of $\beta$ for  the R2D pulse and the analytical approximation, for different values of $\alpha_{12}$. Once again, the analytical expression can accurately predict the values of  $\beta$.}
    \label{fig:R2D_cal}
\end{figure*}

\subsection{Parameter optimization for recursive pulses} \label{r2d_num}

The same procedure can be applied to R2D. We replace $\alpha$ with 3 constants, $\alpha_{12}$, $\alpha_{02}$ and $\alpha_{13}$, each allowing for higher order corrections in the 1-2, 0-2 and 1-3 transitions, respectively. The introduction of these constants correspond to replacing $\alpha$ in Eq.~(\ref{eq:framealpha}) with $\alpha_{12}$ and changing the frame transformations: $\hat{S}_3 \rightarrow \alpha_{02} \hat{S}_3$ and $\hat{S}_7 \rightarrow \alpha_{13} \hat{S}_7$. This changes our pulses to:

\begin{equation}
\label{eq:recursion13}
    \Omega_1(t) = \sqrt{\Omega_2^2+\frac{2\alpha_{13}}{\Delta_3^2}(\dot{\Omega}^2_2+\ddot{\Omega}_2\Omega_2)}
\end{equation}

\begin{equation}
    \Omega_x(t)=\sqrt{\Omega_1^2+\frac{2\alpha_{02}}{\Delta_2^2}(\dot{\Omega}_1^2+\ddot{\Omega}_1\Omega_1)}.
    \label{eq:R1Dopt}
\end{equation}
and \begin{eqnarray}
\label{eq:r2d_num}
\Omega \rightarrow \beta (\Omega_x+\alpha_{12} \Omega_y),\quad \delta' \rightarrow \delta_c.
\end{eqnarray}
with $\Omega_2(t)=\sin^4{\pi t/T}$. To second order in the error, $O({\Omega_x^2}/{\Delta_2^2})$, these parameters will be orthogonal and are free to eliminate higher order errors.

{We plot the infidelity of such a calibrated pulse, termed R2D N in Fig. \ref{fig:fig1_inks} c)} as a function of time. As can be seen, we achieve significant improvements in fidelity, with a fidelity of $99.999$ \% achieved at T$\approx$ 6.75~ns, demonstrating once again the capacity of correcting for higher order terms through calibrating constants. Contrary to linear DRAG, the fidelity stays above this threshold for all times beyond.

Unfortunately, analytical prediction of the three DRAG prefactors becomes intractable, as deviations in any one parameter feed back into the others through higher-order corrections, resulting in a coupled set of nonlinear equations.
Nevertheless, we find a good agreement between our simulations and the experimental results obtained in \cite{GAO2025}.
The parameters can be found in Fig.~\ref{fig:all_alphas}. In contrast to $\alpha_{12}$ and $\alpha_{02}$, $\alpha_{13}$ becomes bigger as time goes on because the maximum amplitude of the pulse becomes smaller with increasing $\alpha_{13}$. This reduces all sources of error at higher orders (and therefore compensates the increase in the 2 photon leakage from $\ket{1} \rightleftarrows\ket{3}$).

For the amplitude and constant detuning, Eqs. (\ref{eq:cnt_det_comp}) and (\ref{eq:beta_3}) remain valid to the same order in O($\Omega_x/\Delta$), with the replacement rule $\alpha \rightarrow \alpha_{12}$, as can be seen in Fig.~\ref{fig:R2D_cal} (a) and (b). Interestingly enough, this means that the correction to the amplitude of the pulse and the constant detuning are mostly defined by $\alpha_{12}$.

One final note regarding the optimization of R2D pulses --  $\alpha_{02}$ and $\alpha_{13}$ will change the minimum gate time. In general, a numerical verification is necessary to extract a minimum gate time. However, if $\alpha_{02}>\alpha_{13}$, the expression for minimum gate time can be reduced to:

\begin{equation}
    T_{min}=\frac{\pi\sqrt{n(\alpha_{13}\Delta_2^2+\alpha_{02}\Delta_3^2)+\nu}}{\Delta_2\Delta_3},
\end{equation}
with
\begin{equation}
    \nu= \sqrt{n^2(\alpha_{13}^2\Delta_2^4+\alpha_{02}^2\Delta_3^4)+2(2-5n)n\alpha_{13}\alpha_{02}\Delta_2^2\Delta_3^2}.
\end{equation}

\section{Dissipative dynamics}\label{dissipation}
\begin{figure}[!t]
    \centering
    \includegraphics[width=1\linewidth]{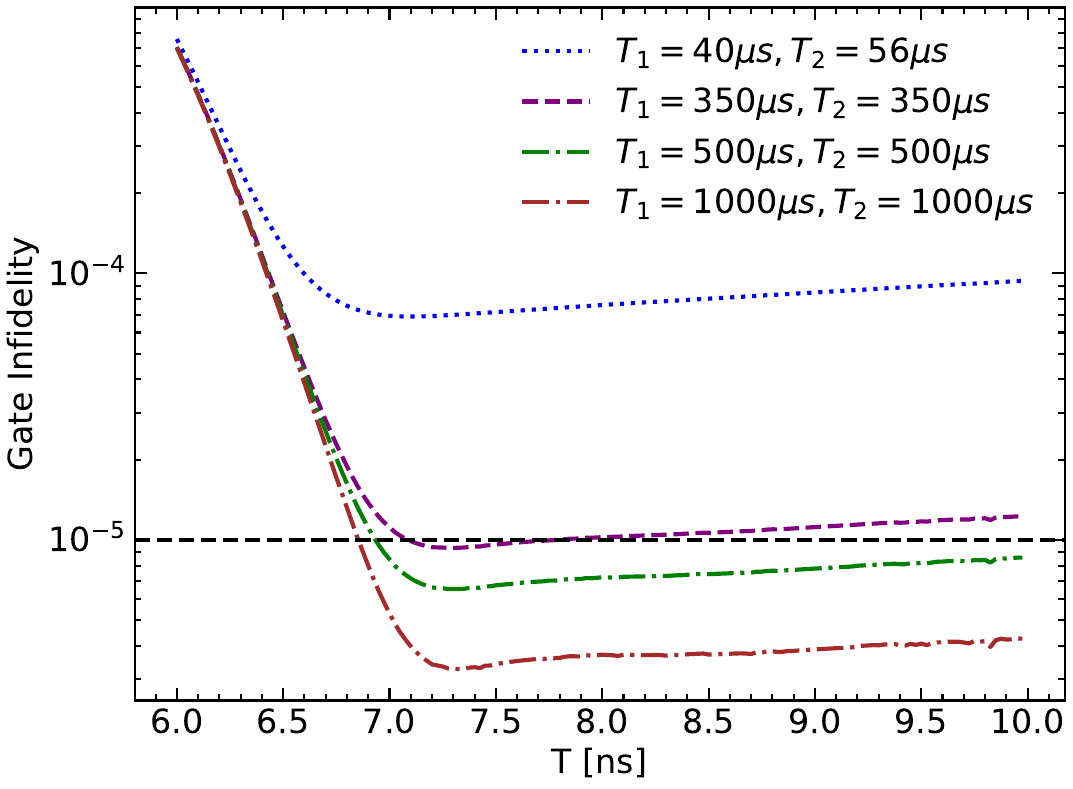}
    \caption{ \textbf{Gate averaged infidelities for a $\pi$ gate using optimized parameters under decoherence}. As $T_1$ and {$T_2$} improves, R2D pulses manage to reach 99.999\% fidelity. }
    \label{fig:deco}
\end{figure}
We calculate the system dynamics considering loss mechanisms including energy relaxation and dephasing. The evolution is described by the master equation in Lindblad form~\cite{Manzano_2020}
\begin{equation}
\label{ME}
    \dot{\rho}(t) = -i[\mathcal{H}(t), \rho(t)] +\gamma \mathcal{D}[\hat{a}]\rho(t) +\gamma_{\phi} \mathcal{D}[\hat{a}^{\dag}\hat{a}]\rho(t),
\end{equation}
where $\rho(t)$ is the density matrix of the system, $\mathcal{D}[\hat{O}]\bullet=O\bullet O^{\dag} - \{O^{\dag}O,\bullet\}/2$ is the Lindbladian super-operator and $\gamma=1/T_{1}$ and {$\gamma_{\phi}=1/T_{2}-1/(2T_1)$ }correspond to the decay rates for energy losses and dephasing, respectively. We have choosen {3 sets of $T_1$ and $T_2$, respectively (40,56) $\mu$s, (350, 350) $\mu$s and (500,500) $\mu$s} , representing a range of  values achievable today \cite{IBM_Machine,wang2022towards}. We also add a fourth set, (1000,1000) $\mu$s, values which we hope will be reached in the near future. Finally, $\hat{a}$ is the annihilation operator describing the transmon system. As the dissipative dynamics are state-dependent instead of unitary-dependent, we need to evaluate the gate performance for all the states of the Bloch sphere of the two-level system and average them~\cite{BOWDREY2002258}.

As can be seen in Fig. \ref{fig:deco}, R2D is capable of achieving a fidelity of 99.999\% even under decoherence, for a gate time of $T=$ 6.88~ns, with a minimum in infidelity at $T=$ 7.33~ns. This demonstrates that single qubit gates with infidelities below $10^{-5}$ are possible in today's hardware.

\section{Summary and Conclusion}\label{Conclu}

In this work, we expand on the DRAG framework by including higher levels not usually considered under weak driving and we show that these levels have a real impact on gate fidelity through multi-photon processes. Taking these in consideration, we extract new analytical expressions for pulses that show that is possible to implement high-fidelity single-qubit gates at faster speeds and stronger driving strengths than previously considered.

We also explore numerical optimization of DRAG and R2D pulses. For the DRAG pulses we obtain theoretical approximations for the calibrating parameters, expressions that we hope will simplify and speed up normal calibration procedures. The obtained expressions for constant detuning are experimentally validated in \cite{GAO2025}. For R2D pulses we show that a moderate optimization allows us to achieve high fidelity pulses faster, even under the effects of system decoherence, a critical step in enabling future quantum computation.

Experimental implementation of this theoretical framework has since been achieved by Y. Gao et al.  \cite{GAO2025}, where R2D pulses were implemented and used to implement single qubit gates with suppressed two-photon leakage channels at world record speeds (6.8 ns) and drive strengths, $\Omega_x/\Delta_2$, whilst maintaining a fidelity of 99.98\%. {This gate is mainly limited by decoherence. The difference between the measured value and the simulated is a consequence of heating of the chip during the action of the gate, leading to a large thermal population ($n\approx0.1$), and the usage of 1 ns of padding time to account for pulse distortion effects. Given that these parameters are chip dependent they were not included in the simulation of Fig. \ref{fig:deco}}. 

Although our equations were motivated by the transmon model, R2D is applicable for any ladder system modelled under the rotating wave approximation~\cite{PhysRevA.88.062318}. Future work is required to go beyond this approximation to remove the remaining three-photon processes in the system,  including the $\ket{1}-\ket{4}$ that will require an inclusion of the $\ket{4}$ level. Moreover, going to shorter times and gate times below 5~ns requires better accounting of waveform generation electronics where distortion can be significant, meanwhile the present optimization procedure for R2D may find negative values inside the square root \cite{GAO2025}. More generally, one may derive higher order and multi-photon corrections to similar strong driving dynamics in entangling gates \cite{kirchhoff2025correction}.

\section*{Acknowledgments}
We thank Manuel Guatto, Dimitrios Georgiadis, Shahrukh Chishti and Asier Galicia for valuable discussions. This work was funded by the Federal Ministry of Education and Research (BMBF) within the framework programme "Quantum technologies - from basic research to market" (Project QSolid, Grant No. 13N16149) and by Horizon Europe program via project QCFD (101080085, HORIZON-CL4-2021-DIGITAL-EMERGING02-10), project OpenSuperQPlus100 (101113946, HORIZON-CL4-2022-QUANTUM-01-SGA) and by the Cluster of Excellence Matter and Light for Quantum Computing (ML4Q2) EXC 2004/2 – 390534769. 
All the data used to generate the figures in this paper can be found in \cite{Repo}.

\appendix

\section{Charge basis Hamiltonian}\label{derivationH}
We start by deriving the effective Hamiltonian of a transmon circuit being driven on resonance with the $\ket{0} \rightleftarrows \ket{1}$ transition. The transmon circuit oscillator is described by the Hamiltonian~\cite{PhysRevA.76.042319}
\begin{equation}
    \hat{\mathcal{H}} = 4E_{C}[\hat{n}-n_g(t)]^2-E_{J}\cos(\hat{\varphi})
\label{eq:transmon hamiltonian exact}
\end{equation}
where $E_{C}$ and $E_{J}$ represent the charge and Josephson energies, and $n_g(t)$ is the dimensionless gate voltage. The operator $\hat{n}$ is the charge operator, indicating the number of Cooper pairs, and $\hat{\varphi}$ denotes the phase operator. We can implement single-qubit operations by driving the circuit with an external source using $n_g(t)=n_{0}(t)\cos(\omega_{d}t)$ resulting in the Hamiltonian
\begin{eqnarray}
    \hat{H}_{{\rm{ctrl}}}=\Omega(t)\cos(\omega_{d}t)\hat{n}
\end{eqnarray}
where $\Omega(t)=-8E_Cn_{0}(t)$ is the complex drive envelope, and $\omega_{d}$ is the drive frequency. For the low-lying energy spectrum of this device, we can model it as an approximate Duffing oscillator~\cite{Khani2009Optimal,Blais2021Circuit}, where the controls now are given by $\hat{n}\propto(\hat{b}^{\dag}+\hat{b})$, where $\hat{b}$ is the annihilation operator of a linear oscillator. For the free energy term, we expand the cosine up to fourth order leading to  
\begin{equation}
    \hat{H}_0^{\textnormal{duf}} = \omega_q \hat{b}^{\dagger}\hat{b}+\frac{\Delta}{2}\hat{b}^{\dagger}\hat{b}^{\dagger}\hat{b}\hat{b}
    ,
\end{equation}
with $\omega_q = \sqrt{8 E_J E_C} - E_C$ and $\Delta=-E_C$. However, the eigenenergies and coupling strengths deviate from the Duffing model at higher levels due to the higher-order expansion of the cosine potential~\cite{Khani2009Optimal}. The accurate description of the transmon circuit requires us to diagonalize $\hat{\mathcal{H}}$ up to a truncation level $N_{\rm{max}}$, which gives
\begin{equation}
    H_0 = \sum_{k=0}^{N_{\rm{max}}} \omega_k \ket{k}\bra{k},
\end{equation}
whereas the charge operator takes the form
\begin{eqnarray}\nonumber
    \hat{n}&=&\sum_{k=0}^{N_{{\rm{max}}}-1}\bigg[\lambda_{k,k+1}\ketbra{k}{k+1}\\
            &+&\sum_{j=1} \lambda_{k,k+2j+1}\ketbra{k}{k+2j+1}\bigg]+\rm{h.c}.
\end{eqnarray}
where $\lambda$ denotes the corresponding matrix element.
The maximal $j$ is chosen such that $k+2j+1$ falls within the truncated levels. Unlike the Duffing oscillator model, the charge operator contains additional transitions corresponding to the underlying parity symmetry from the Mathieu functions. One can show that the coupling terms beyond the ladder structure are orders of magnitude smaller and are suppressed by the rotating-wave approximation.

From this point, it is more convenient to express the Hamiltonian in the rotating frame defined by the transformation $R = \exp\left(-i\omega_d t\sum_{k}k\ketbra{k}{k}\right)$, where $\omega_d$ is the selected driving transition frequency. This leads to the following total Hamiltonian:
\begin{align}
\label{HT}
    \hat{H}=&\sum_{k=0}^{N_{{\rm{max}}}}\tilde{\Delta}_{k}\ket{k}\bra{k}+\Omega(t)\cos(\omega_d t)
    \Biggr[\nonumber\\
    &\sum_{k=0}^{N_{\rm{max}}-1}\sum_{j=1}\lambda_{k,k+2j+1}\ketbra{k}{k+2j+1}e^{-(2j+1)i\omega_d t}
    \nonumber\\
    &+
    \sum_{k=0}^{N_{{\rm{max}}}-1}\lambda_{k,k+1}\ketbra{k}{k+1}e^{-i\omega_d t}
    +\rm{h.c}
    \Biggr]
    ,
\end{align}
where $\Delta_{k}=\omega_{k}-k\omega_{d}$ is the detuning between the $k$-th level with the $k$-th driving frequency harmonic. In this frame, all coupling terms, except for $\ket{k}\leftrightarrow\ket{k+1}$, and all the counter-rotating components oscillate rapidly. Therefore, we can neglect those small contributions, leading to the Hamiltonian in Eq.~(\ref{Exp:RWA}).

\section{Fidelity Expressions}\label{Fid_metric}
In this appendix, we define what are the relevant metric to benchmark the performance of our single-qubit gate. In the case of dynamics without dissipative effects, the gate infidelity is defined as $\mathcal{E}(\hat{\mathcal{U}}_Q)=1-\mathcal{F} [\hat{\mathcal{U}}_Q]$ where $\mathcal{F} [\hat{\mathcal{U}}_Q]$ is the gate fidelity defined as
\begin{eqnarray}
\mathcal{F}[\hat{\mathcal{U}}_Q]&=&\frac{\Tr [\hat{\mathcal{U}}_Q \hat{\mathcal{U}}^{\dagger}_Q]}{d(d+1)}
    + \frac{ \left|\Tr [\hat{\mathcal{U}}_Q \hat{\mathcal{U}}_I^{\dagger}] \right|^2}{d(d+1)},
\end{eqnarray}
where $\hat{\mathcal{U}}_Q$ is the time evolution operator at the final time projected in the computational subspace and $\hat{\mathcal{U}}_I=\hat{\sigma}_{0,1}^{x}$ is the target $\pi$-rotation.

In presence of dissipative effects, the metric must change because the master equation strongly depends on the initial state of the system, requiring sampling to compute the gate error. Nevertheless, it is possible avoid sampling by solving the master equation for different initial states given by the poles of the Bloch sphere spanned by the eigenstates of the Pauli matrices projected in the qubit subspace $\{\hat{\sigma}^{x}_{0,1},\hat{\sigma}^{y}_{0,1},\hat{\sigma}^{z}_{0,1}\}$

\begin{eqnarray}
\mathcal{F}_{d}[\hat{\mathcal{U}}_Q]&=&\frac{1}{6}\sum_{j=\pm,\pm x, \pm y}{\rm{Tr}}[\hat{\mathcal{U}}_Q\rho_{j}\hat{\mathcal{U}}_Q^{\dag}\mathcal{M}(\rho_j)],
\end{eqnarray}
where $\rho_j$ are the eigenstates of the Pauli matrices projected in the qubit subspace $\{\sigma^{x}_{0,1},\sigma^{y}_{0,1},\sigma^{z}_{0,1}\}$, moreover, $\mathcal{M}(\rho_j)$ is the state obtained by solving Eq.~(\ref{ME}) considering $\rho_j$ as its intial state.

\section{Third order Hamiltonian and corrections}\label{supralinear_corrections_annex}

Expanding $\hat{\mathcal{H}}_4(t)$[Eq.~(\ref{eq:lastH})]  up to third order in $\Omega_x/ \Delta_2$ and keeping only the qubit subspace and its transitions we obtain:

\begin{eqnarray}
    \hat{\mathcal{H}}_5(t) &=& \bigg[\frac{\Omega_x}{2}+\mathcal{E}^x_{01}\Omega_x^3\bigg]\hat{\sigma}_{0,1}^x+\Omega_x(\mathcal{E}^{y1}_{12}\Omega_1\dot\Omega_1+\mathcal{E}^{y2}_{12}\Omega_2\dot\Omega_2)\hat{\sigma}_{1,2}^y \nonumber \\
    &&+ \Omega_x(\mathcal{E}^{x0}_{12}\Omega_x^2+\mathcal{E}^{x1}_{12}\Omega_1^2+\mathcal{E}^{x2}_{12}\Omega_2^2)\hat{\sigma}_{1,2}^x \nonumber \\
    &&+ \Omega_x(\mathcal{E}^{y1}_{03}\Omega_1\dot{\Omega}_1+ \mathcal{E}^{y2}_{03}\Omega_2\dot{\Omega}_2)\sigma_{0,3}^y \nonumber \\
    &&
    + \Omega_x(\mathcal{E}^{x0}_{03}\Omega_x^2+\mathcal{E}^{x1}_{03}\Omega_1^2+\mathcal{E}^{x2}_{03}\Omega_2^2)\sigma_{0,3}^x + O(\Omega^4)
    \label{eq:finalH}
\end{eqnarray}

The individual constants are given by
\begin{eqnarray}
    \mathcal{E}^x_{01} &=& \frac{(4-\lambda_2^2)}{16\Delta_2^2} \\
    \mathcal{E}^{y1}_{12}&=&-\frac{\lambda_2\lambda_3^2(\Delta_2^2-\Delta_3^2)(2\Delta_2-\Delta_3)^2}{8\Delta_2^5\Delta_3^2}  ,\\
    \mathcal{E}^{y2}_{12}&=&-\frac{\lambda_2(\Delta_2^2+(2\Delta_2-\Delta_3)^2\lambda_3^2)}{8\Delta_2^5} ,\\
    \mathcal{E}^{x0}_{12}&=& \frac{\lambda_2(-3\Delta_3\lambda_3^2+\Delta_2(31-14\lambda_2^2+10\lambda_3^2)}{48\Delta_2^3} ,
    \end{eqnarray}
\begin{eqnarray}
    \mathcal{E}^{x1}_{12}&=& \frac{\lambda_2(\Delta_2^3+\Delta_3\lambda_3^2(2\Delta_2-\Delta_3)^2)}{16\Delta_2^5} ,\\
    \mathcal{E}^{x2}_{12}&=& \frac{\lambda_2\lambda_3^2(\Delta_2^2-\Delta_3^2)(2\Delta_2-\Delta_3)^2}{16\Delta_2^5\Delta_3},\\
    \mathcal{E}^{y1}_{03}&=& \frac{\lambda_2\lambda_3(2\Delta_2-\Delta_3)}{4\Delta_2^4} ,\\
    \mathcal{E}^{y2}_{03}&=&\frac{\lambda_2\lambda_3(2\Delta_2-\Delta_3)(\Delta_2^2-\Delta_3^2)}{8\Delta_2^4\Delta_3} ,\\
    \mathcal{E}^{x0}_{03}&=&\frac{\lambda_2\lambda_3(3\Delta_2-\Delta_3)}{48\Delta_2^3} ,\\
    \mathcal{E}^{x1}_{03}&=& \frac{\lambda_2\lambda_3(\Delta_3-2\Delta_2)(\Delta_2+\Delta_3)}{16\Delta_2^4},\\
    \mathcal{E}^{x2}_{03}&=& \frac{\lambda_2\lambda_3(\Delta_3-2\Delta_2)(\Delta_2^2-\Delta_3^2)}{16\Delta_2^4\Delta_3} .
\end{eqnarray}
 To correct higher order terms in $\sigma^x_{0,1}$ we use the linear shift in Eq. \ref{eq:Cube}. To correct the higher order terms in $\sigma^x_{1,2}$ and $\sigma^y_{1,2}$ we follow the transformation generated by Eq. \ref{eq:Agenerator} and Eq. \ref{eq:Aswitch} introducing an auxiliary variable defined as:

\begin{eqnarray}
A_{0,1}(t)&=&\nonumber\frac{2\Delta_2^2}{\lambda_2}\big{(}\Omega_x(\mathcal{E}^{x0}_{12}\Omega_x^2+\mathcal{E}^{x1}_{12}\Omega_1^2+\mathcal{E}^{x2}_{12}\Omega_2^2)\hat{\sigma}_{1,2}^x \\ \nonumber
&-& \dot{\Omega}_x(\mathcal{E}^{y1}_{12}\Omega_1\dot\Omega_1+\mathcal{E}^{y2}_{12}\Omega_2\dot\Omega_2) \\ \nonumber
&-&  \Omega_x(\mathcal{E}^{y1}_{12}\Omega_1\dot\Omega_1^2+\mathcal{E}^{y2}_{12}\Omega_2\dot\Omega_2^2) \\ 
&-& \Omega_x(\mathcal{E}^{y1}_{12}\Omega_1\ddot\Omega_1+\mathcal{E}^{y2}_{12}\Omega_2\ddot\Omega_2)\big{)}
\label{eq:Adef}
\end{eqnarray}

\section{Magnus Expansion of Toggling Frame}
\label{Magnus_App}

Given $\mathcal{H}_{tog}$, where we have with $\mathcal{E}_{11}^{(\alpha)}=\frac{(\alpha-1)\alpha\lambda_2^2}{8\Delta_2^3}$, $\mathcal{E}^{x0}_{01}=\frac{4\alpha^2+\lambda_2^2(1-2\alpha)}{8\Delta_2^2}$, and  $\mathcal{E}^{x1}_{01}=\frac{\alpha}{\Delta_2}$,{ in order} to obtain $\beta$ and $\delta_c$ we have to compute the computational subspace of $\mathcal{\hat{Z}}$, {where the element $|i\rangle\langle j|$ of $\mathcal{\hat{Z}}$ is represented by $\mathcal{\hat{Z}}$[i,j] = $\int_{0}^{T}dt \langle i|\hat{\mathcal{H}}_{\rm{tog}}(t)|j \rangle$}:

\begin{equation}
   \hat{\mathcal{Z}}_{1}[1,1]  = i\int_0^T \left(\left[\delta_c+\mathcal{E}_{11}\Omega_x^2\right]\cos{\theta}+\mathcal{E}_{11}^{(\alpha)}\Omega_x\dot{\Omega}_x\sin{\theta}\right)dt
    \label{eq:err11}
\end{equation}

\begin{eqnarray}
    \hat{\mathcal{Z}}_{1}[0,1] &=& \frac{1}{2}\int_0^T dt\bigg(\Omega_x-\dot\theta+i\left[\delta_c+\mathcal{E}_{11}\Omega_x^2\right]\sin{\theta} \nonumber\\ &&- \mathcal{E}^{x0}_{01}\Omega_x^3+\mathcal{E}^{x1}_{01}\Omega_x\delta_c+i\mathcal{E}_{11}^{(\alpha)}\Omega_x\dot{\Omega}_x\cos{\theta}\bigg),\nonumber \\
    \label{eq:err01}
\end{eqnarray}

 Note that both $\Omega_x$ and $\theta$ are time-dependent functions in the above equations.

Given the rotation angle $\theta=\int \frac{2\pi}{T}\sin^2({\pi \tau/T})d\tau$, that smoothly increases from 0 to $\pi$, then $\cos{\theta}$ will be an odd function over the period of integration $[0,T]$ around $T/2$ (it corresponds to cos([0,$\pi$]) which is odd). This simplifies the first term in Eq.~(\ref{eq:err11}) to zero -- we only have products of constants and odd functions or even functions and odd functions. The same logic applies to the second component where now $\sin{\theta}$ is even but $\dot{\Omega}_x$
is odd. This way we can use the constant detuning $\delta_c$ and the amplitude of the pulse to make Eq. (\ref{eq:err11}) $=0$. It is important to note that this is only true for the $\pi$ rotation.

 In order to compute the analytical DRAG prefactor, we need the elements of the Magnus expansion of the Toggling frame of transitions $\ket{1}\rightleftarrows\ket{2}$ and $\ket{0}\rightleftarrows\ket{2}$. To the second order, these are:

\begin{widetext}

\begin{equation}
    \hat{\mathcal{Z}}_{1}[1,2]= \int^{T}_{0}\frac{\lambda_2 e^{-i\phi_1}}{4\Delta_2}\big{(}i( \Omega_x^2\sin{\frac{\theta}{2}-2\dot{\Omega}_x}\cos{\frac{\theta}{2}})-\alpha \frac{-4i\Delta_2\dot{\Omega}_x\cos{\frac{\theta}{2}}+\Omega_x(2\dot{\Omega}_x+i\Delta_2\Omega_x)\sin{\frac{\theta}{2}}}{2\Delta_2}+\alpha^2\frac{\Omega_x\dot{\Omega}_x\sin{\frac{\theta}{2}}}{\Delta_2}\big{)}dt
    \label{eq:Z12}
\end{equation}
and
\begin{equation}
    \hat{\mathcal{Z}}_{1}[0,2] =\int^{T}_{0} \frac{\lambda_2 e^{-i\phi_1}}{4\Delta_2}\big{(}( \Omega_x^2\cos{\frac{\theta}{2}+2\dot{\Omega}_x}\sin{\frac{\theta}{2}})-\alpha \frac{4\Delta_2\dot{\Omega}_x\sin{\frac{\theta}{2}}+\Omega_x(-2i\dot{\Omega}_x+\Delta_2\Omega_x)\cos{\frac{\theta}{2}}}{2\Delta_2}-i\alpha^2\frac{\Omega_x\dot{\Omega}_x\cos{\frac{\theta}{2}}}{\Delta_2}\big{)}dt
    \label{eq:Z02}
\end{equation}
    
\end{widetext}
with $\hat{\mathcal{Z}}_{1}[2,1] =\Bar{\hat{\mathcal{Z}}}_{1}[1,2] $ and $\hat{\mathcal{Z}}_{1}[2,0] =\Bar{\hat{\mathcal{Z}}}_{1}[0,2] $.

To obtain \ref{eq:approx_alphas}, we define:
\begin{eqnarray}\nonumber
    x_1&=&(h_{12}^{1}+2h_{12}^{2})(\Bar{h}_{12}^{0}+\Bar{h}_{12}^{1}+\Bar{h}_{12}^{2})\\\nonumber
    &+&(h_{02}^{1}+2h_{02}^{2})(\Bar{h}_{02}^{0}+\Bar{h}_{02}^{1}+\Bar{h}_{02}^{2}),\\\nonumber
        x_2&=&h_{12}^{1}\Bar{h}_{12}^1+2h_{12}^{1}\Bar{h}_{12}^2+2h_{12}^{2}(\Bar{h}_{12}^{0}+2\Bar{h}_{12}^{1}    +3\Bar{h}_{12}^{2})\nonumber\\
        &+&h_{02}^{1}\Bar{h}_{02}^1+2h_{02}^{1}\Bar{h}_{02}^2+2h_{02}^{2}(\Bar{h}_{02}^{0}+2\Bar{h}_{02}^{1}++3\Bar{h}_{02}^{2}),\nonumber
\end{eqnarray}
where the symbol $h_{ij}^{k}$ represents the integral from 0 to $T$ of the $k$th $\alpha$ power in $\hat{\mathcal{H}}_{{\rm{tog}}}[i,j]$, i.e $\hat{\mathcal{Z}}_{1}[1,2]= \int_0^T \hat{\mathcal{H}}_{{\rm{tog}}}[1,2]dt = h_{12}^{0}+h_{12}^{1}\alpha+h_{12}^2\alpha^2$ and $\Bar{x}_1$ ($\Bar{x}_2$) is the complex conjugate of $x_1$ ($x_2$).

\section{R1D prefactors optimization}\label{r1d_num}

\begin{figure}[!h]
    \centering
    \includegraphics[width=0.85\linewidth]{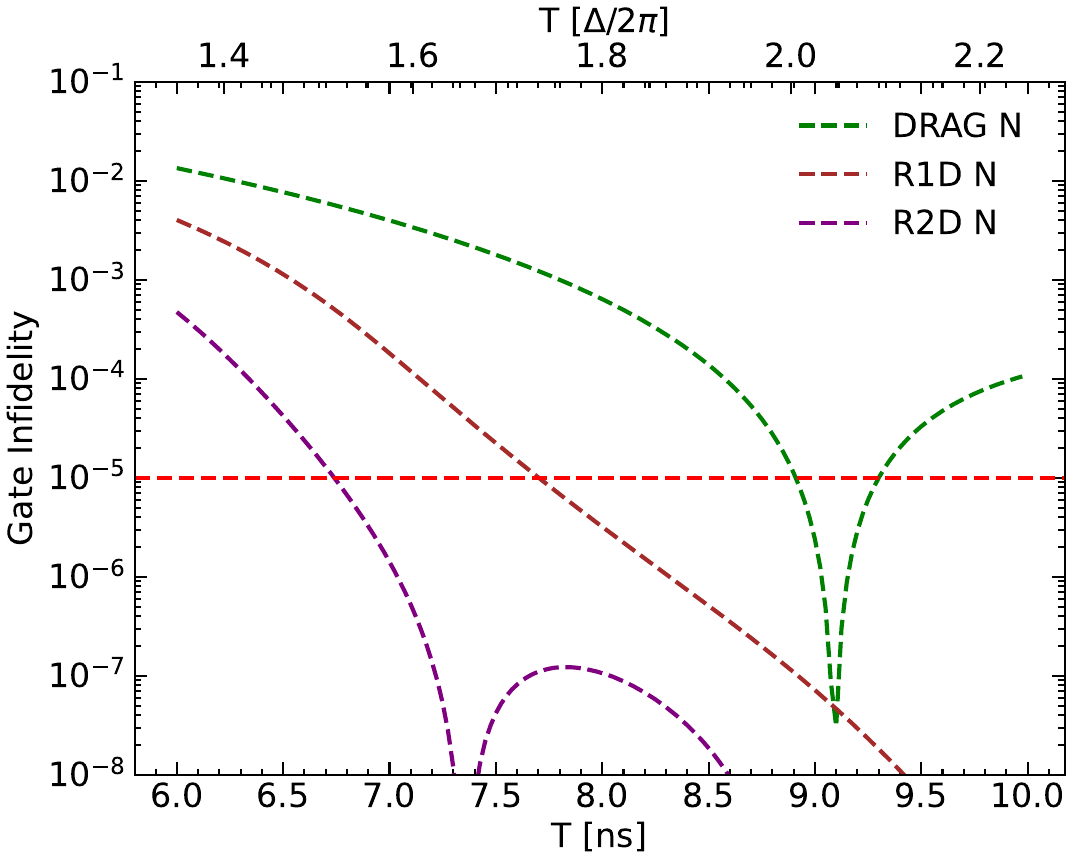}
    \caption{Gate Infidelities for calibrated linear DRAG pulses (green), calibrated R1D pulses (brown) and calibrated R2D pulses (purple). Although R1D performs better than linear DRAG, it fails to achieve fidelities above 99.999 \% for gate times shorter than $T\approx$ 7.7 ~ns}
    \label{fig:inf_R1D}
\end{figure}

\begin{figure}[!b]
    \centering
    \includegraphics[width=0.85\linewidth]{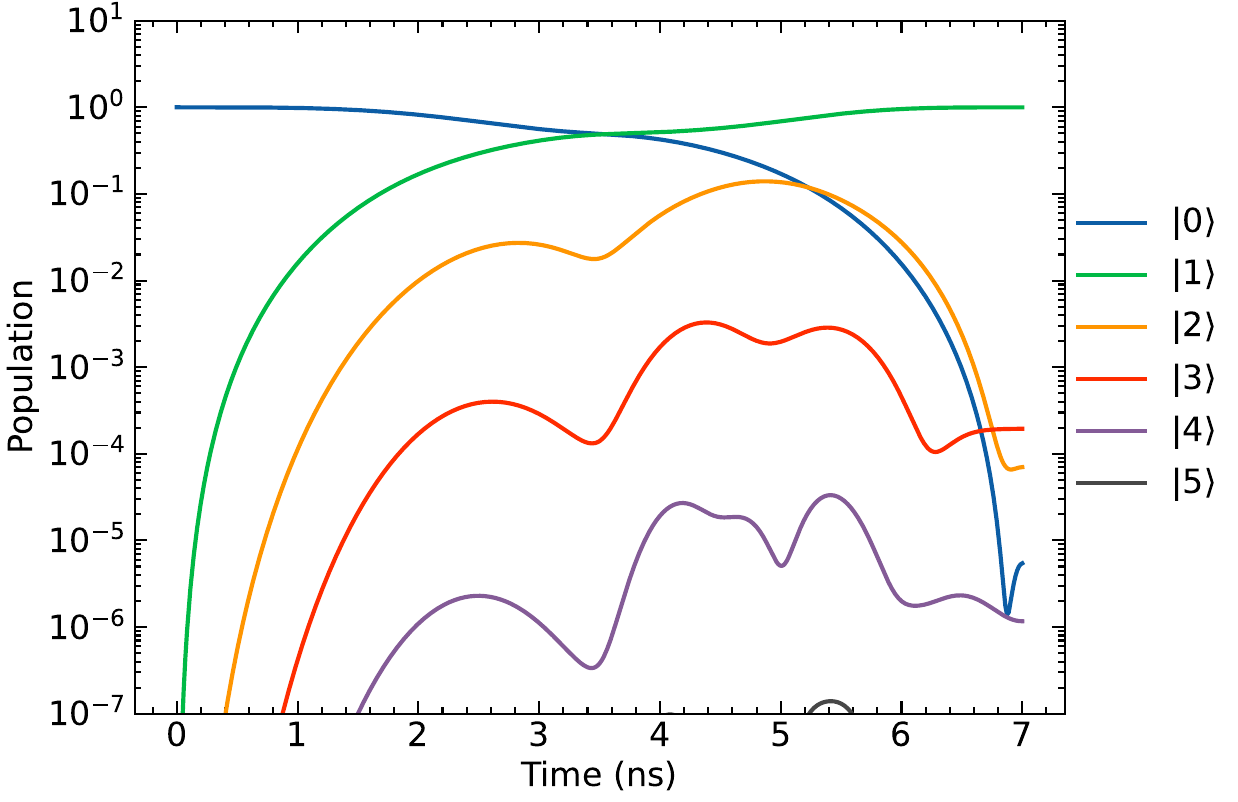}
    \caption{Evolution of population for a $\pi$ gate using a calibrated R1D pulse. As can be seen, leakage to $|3\rangle$ is substantial and hampers gate performance}
    \label{fig:pop_evo_R1d}
\end{figure}

We can also optimize R1D pulses by using Eq. \ref{eq:R1Dopt} and starting with $\Omega_1=\sin^3{\pi t/T}$. As can be seen in Fig. \ref{fig:inf_R1D}  although we see an improvement over optimized linear DRAG, R1D does not match R2D. Plotting population evolution for a gate time of T=7 ~[ns] in Fig. \ref{fig:pop_evo_R1d} explains why -  we see that leakage to $|3\rangle$ clearly limits our gate performance. This lends even more strength to the need for R2D pulses.

\section{Ansatz\"e Selection}
Here, we will show how to obtain the pulse from Eq.~(\ref{eq:bl_pulse}). Our starting point is the boundary conditions given in Eq~(\ref{Exp2}) together with the additional one provided by both R1D and R2D, thus, we can choose a Fourier ansatz of the form

\begin{equation}
    \Omega(t) = \frac{1}{2} +\frac{1}{2(k-1)}\bigg[\cos\bigg[j \frac{2\pi}{T}t\bigg]-k\cos\bigg[n \frac{2\pi}{T}t\bigg]\bigg],
\end{equation}
this functional form ensures that the first and third derivatives are zero at $t=0$ and $t=T$, then, we need to find the condition for the second derivative at both times, obtaining that the pulse from Eq.~(\ref{eq:bl_pulse}) achieves the lower error for a gate faster than ten nanoseconds.
\begin{equation}
    \ddot{\Omega}=-\bigg[\frac{2\pi}{T}\bigg]^2\bigg[j^2\cos\bigg[j \frac{2\pi}{T}t\bigg] -kn^2\cos\bigg[n \frac{2\pi}{T}t\bigg]\bigg].
\end{equation}
To obtain null derivative at $t=0$ and $t=T$, we obtain the relation
\begin{equation}
    k = \frac{j^2}{n^2},\quad \forall k\neq 1.
\end{equation}
Table~\ref{tab:my_label} shows the optimal gate time $T$ to obtain gate error bellow $10^{-4}$ for different value of the triad $\{n,j,k\}$. Obtaining that 
\begin{table}[b!]
    \centering
    \setlength{\tabcolsep}{10pt}
    \renewcommand{\arraystretch}{1.2}
    \begin{tabular}{|c|c|c|c|}
        \hline
        $n$ & $j$ & $k$ & $T_{99.99}~$(ns) \\
        \hline
        1 & 2 & 4 & 10.43 \\\hline
        \rowcolor{green!20}
        \textbf{1} & \textbf{3} & \textbf{9} & \textbf{\cellcolor{green!40}8.42} \\
        \hline
        1 & 4 & 16 & 8.45 \\\hline
        1 & 5 & 25 & 8.60 \\\hline
        2 & 3 & $9/4$ & 18.0 \\\hline
        2 & 4 & 8 & 18.0 \\\hline
        2 & 5 & $25/4$ & 18.0 \\
        \hline
    \end{tabular}
    \caption{Optimal gate times $T$ achieving an error below $10^{-4}$ for different choices of the ansatz parameters $\{n, j, k\}$. The highlighted entry corresponds to the fastest configuration.}
    \label{tab:my_label}
\end{table}

\bibliographystyle{apsrev4-2}
\bibliography{References}

\begin{thebibliography}{36}%
\makeatletter
\providecommand \@ifxundefined [1]{%
 \@ifx{#1\undefined}
}%
\providecommand \@ifnum [1]{%
 \ifnum #1\expandafter \@firstoftwo
 \else \expandafter \@secondoftwo
 \fi
}%
\providecommand \@ifx [1]{%
 \ifx #1\expandafter \@firstoftwo
 \else \expandafter \@secondoftwo
 \fi
}%
\providecommand \natexlab [1]{#1}%
\providecommand \enquote  [1]{``#1''}%
\providecommand \bibnamefont  [1]{#1}%
\providecommand \bibfnamefont [1]{#1}%
\providecommand \citenamefont [1]{#1}%
\providecommand \href@noop [0]{\@secondoftwo}%
\providecommand \href [0]{\begingroup \@sanitize@url \@href}%
\providecommand \@href[1]{\@@startlink{#1}\@@href}%
\providecommand \@@href[1]{\endgroup#1\@@endlink}%
\providecommand \@sanitize@url [0]{\catcode `\\12\catcode `\$12\catcode
  `\&12\catcode `\#12\catcode `\^12\catcode `\_12\catcode `\%12\relax}%
\providecommand \@@startlink[1]{}%
\providecommand \@@endlink[0]{}%
\providecommand \url  [0]{\begingroup\@sanitize@url \@url }%
\providecommand \@url [1]{\endgroup\@href {#1}{\urlprefix }}%
\providecommand \urlprefix  [0]{URL }%
\providecommand \Eprint [0]{\href }%
\providecommand \doibase [0]{https://doi.org/}%
\providecommand \selectlanguage [0]{\@gobble}%
\providecommand \bibinfo  [0]{\@secondoftwo}%
\providecommand \bibfield  [0]{\@secondoftwo}%
\providecommand \translation [1]{[#1]}%
\providecommand \BibitemOpen [0]{}%
\providecommand \bibitemStop [0]{}%
\providecommand \bibitemNoStop [0]{.\EOS\space}%
\providecommand \EOS [0]{\spacefactor3000\relax}%
\providecommand \BibitemShut  [1]{\csname bibitem#1\endcsname}%
\let\auto@bib@innerbib\@empty
%</preamble>
\bibitem [{\citenamefont {Krantz}\ \emph {et~al.}(2019)\citenamefont {Krantz},
  \citenamefont {Kjaergaard}, \citenamefont {Yan}, \citenamefont {Orlando},
  \citenamefont {Gustavsson},\ and\ \citenamefont {Oliver}}]{Krantz2019}%
  \BibitemOpen
  \bibfield  {author} {\bibinfo {author} {\bibfnamefont {P.}~\bibnamefont
  {Krantz}}, \bibinfo {author} {\bibfnamefont {M.}~\bibnamefont {Kjaergaard}},
  \bibinfo {author} {\bibfnamefont {F.}~\bibnamefont {Yan}}, \bibinfo {author}
  {\bibfnamefont {T.~P.}\ \bibnamefont {Orlando}}, \bibinfo {author}
  {\bibfnamefont {S.}~\bibnamefont {Gustavsson}},\ and\ \bibinfo {author}
  {\bibfnamefont {W.~D.}\ \bibnamefont {Oliver}},\ }\href
  {https://doi.org/10.1063/1.5089550} {\bibfield  {journal} {\bibinfo
  {journal} {Applied Physics Reviews}\ }\textbf {\bibinfo {volume} {6}},\
  \bibinfo {pages} {021318} (\bibinfo {year} {2019})}\BibitemShut {NoStop}%
\bibitem [{\citenamefont {Blais}\ \emph {et~al.}(2021)\citenamefont {Blais},
  \citenamefont {Grimsmo}, \citenamefont {Girvin},\ and\ \citenamefont
  {Wallraff}}]{Blais2021Circuit}%
  \BibitemOpen
  \bibfield  {author} {\bibinfo {author} {\bibfnamefont {A.}~\bibnamefont
  {Blais}}, \bibinfo {author} {\bibfnamefont {A.~L.}\ \bibnamefont {Grimsmo}},
  \bibinfo {author} {\bibfnamefont {S.~M.}\ \bibnamefont {Girvin}},\ and\
  \bibinfo {author} {\bibfnamefont {A.}~\bibnamefont {Wallraff}},\ }\href
  {https://doi.org/10.1103/RevModPhys.93.025005} {\bibfield  {journal}
  {\bibinfo  {journal} {Reviews of Modern Physics}\ }\textbf {\bibinfo {volume}
  {93}},\ \bibinfo {pages} {025005} (\bibinfo {year} {2021})},\ \Eprint
  {https://arxiv.org/abs/2005.12667} {arXiv:2005.12667 [quant-ph]} \BibitemShut
  {NoStop}%
\bibitem [{\citenamefont {Koch}\ \emph {et~al.}(2007)\citenamefont {Koch},
  \citenamefont {Yu}, \citenamefont {Gambetta}, \citenamefont {Houck},
  \citenamefont {Schuster}, \citenamefont {Majer}, \citenamefont {Blais},
  \citenamefont {Devoret}, \citenamefont {Girvin},\ and\ \citenamefont
  {Schoelkopf}}]{PhysRevA.76.042319}%
  \BibitemOpen
  \bibfield  {author} {\bibinfo {author} {\bibfnamefont {J.}~\bibnamefont
  {Koch}}, \bibinfo {author} {\bibfnamefont {T.~M.}\ \bibnamefont {Yu}},
  \bibinfo {author} {\bibfnamefont {J.}~\bibnamefont {Gambetta}}, \bibinfo
  {author} {\bibfnamefont {A.~A.}\ \bibnamefont {Houck}}, \bibinfo {author}
  {\bibfnamefont {D.~I.}\ \bibnamefont {Schuster}}, \bibinfo {author}
  {\bibfnamefont {J.}~\bibnamefont {Majer}}, \bibinfo {author} {\bibfnamefont
  {A.}~\bibnamefont {Blais}}, \bibinfo {author} {\bibfnamefont {M.~H.}\
  \bibnamefont {Devoret}}, \bibinfo {author} {\bibfnamefont {S.~M.}\
  \bibnamefont {Girvin}},\ and\ \bibinfo {author} {\bibfnamefont {R.~J.}\
  \bibnamefont {Schoelkopf}},\ }\href
  {https://doi.org/10.1103/PhysRevA.76.042319} {\bibfield  {journal} {\bibinfo
  {journal} {Phys. Rev. A}\ }\textbf {\bibinfo {volume} {76}},\ \bibinfo
  {pages} {042319} (\bibinfo {year} {2007})}\BibitemShut {NoStop}%
\bibitem [{\citenamefont {Chen}\ \emph {et~al.}(2016)\citenamefont {Chen},
  \citenamefont {Kelly}, \citenamefont {Quintana}, \citenamefont {Barends},
  \citenamefont {Campbell}, \citenamefont {Chen}, \citenamefont {Chiaro},
  \citenamefont {Dunsworth}, \citenamefont {Fowler}, \citenamefont {Lucero},
  \citenamefont {Jeffrey}, \citenamefont {Megrant}, \citenamefont {Mutus},
  \citenamefont {Neeley}, \citenamefont {Neill}, \citenamefont {O'Malley},
  \citenamefont {Roushan}, \citenamefont {Sank}, \citenamefont {Vainsencher},
  \citenamefont {Wenner}, \citenamefont {White}, \citenamefont {Korotkov},\
  and\ \citenamefont {Martinis}}]{PhysRevLett.116.020501}%
  \BibitemOpen
  \bibfield  {author} {\bibinfo {author} {\bibfnamefont {Z.}~\bibnamefont
  {Chen}}, \bibinfo {author} {\bibfnamefont {J.}~\bibnamefont {Kelly}},
  \bibinfo {author} {\bibfnamefont {C.}~\bibnamefont {Quintana}}, \bibinfo
  {author} {\bibfnamefont {R.}~\bibnamefont {Barends}}, \bibinfo {author}
  {\bibfnamefont {B.}~\bibnamefont {Campbell}}, \bibinfo {author}
  {\bibfnamefont {Y.}~\bibnamefont {Chen}}, \bibinfo {author} {\bibfnamefont
  {B.}~\bibnamefont {Chiaro}}, \bibinfo {author} {\bibfnamefont
  {A.}~\bibnamefont {Dunsworth}}, \bibinfo {author} {\bibfnamefont {A.~G.}\
  \bibnamefont {Fowler}}, \bibinfo {author} {\bibfnamefont {E.}~\bibnamefont
  {Lucero}}, \bibinfo {author} {\bibfnamefont {E.}~\bibnamefont {Jeffrey}},
  \bibinfo {author} {\bibfnamefont {A.}~\bibnamefont {Megrant}}, \bibinfo
  {author} {\bibfnamefont {J.}~\bibnamefont {Mutus}}, \bibinfo {author}
  {\bibfnamefont {M.}~\bibnamefont {Neeley}}, \bibinfo {author} {\bibfnamefont
  {C.}~\bibnamefont {Neill}}, \bibinfo {author} {\bibfnamefont {P.~J.~J.}\
  \bibnamefont {O'Malley}}, \bibinfo {author} {\bibfnamefont {P.}~\bibnamefont
  {Roushan}}, \bibinfo {author} {\bibfnamefont {D.}~\bibnamefont {Sank}},
  \bibinfo {author} {\bibfnamefont {A.}~\bibnamefont {Vainsencher}}, \bibinfo
  {author} {\bibfnamefont {J.}~\bibnamefont {Wenner}}, \bibinfo {author}
  {\bibfnamefont {T.~C.}\ \bibnamefont {White}}, \bibinfo {author}
  {\bibfnamefont {A.~N.}\ \bibnamefont {Korotkov}},\ and\ \bibinfo {author}
  {\bibfnamefont {J.~M.}\ \bibnamefont {Martinis}},\ }\href
  {https://doi.org/10.1103/PhysRevLett.116.020501} {\bibfield  {journal}
  {\bibinfo  {journal} {Phys. Rev. Lett.}\ }\textbf {\bibinfo {volume} {116}},\
  \bibinfo {pages} {020501} (\bibinfo {year} {2016})}\BibitemShut {NoStop}%
\bibitem [{\citenamefont {Motzoi}\ \emph {et~al.}(2009)\citenamefont {Motzoi},
  \citenamefont {Gambetta}, \citenamefont {Rebentrost},\ and\ \citenamefont
  {Wilhelm}}]{motzoi2009simple}%
  \BibitemOpen
  \bibfield  {author} {\bibinfo {author} {\bibfnamefont {F.}~\bibnamefont
  {Motzoi}}, \bibinfo {author} {\bibfnamefont {J.~M.}\ \bibnamefont
  {Gambetta}}, \bibinfo {author} {\bibfnamefont {P.}~\bibnamefont
  {Rebentrost}},\ and\ \bibinfo {author} {\bibfnamefont {F.~K.}\ \bibnamefont
  {Wilhelm}},\ }\href {https://doi.org/10.1103/PhysRevLett.103.110501}
  {\bibfield  {journal} {\bibinfo  {journal} {Phys. Rev. Lett.}\ }\textbf
  {\bibinfo {volume} {103}},\ \bibinfo {pages} {110501} (\bibinfo {year}
  {2009})}\BibitemShut {NoStop}%
\bibitem [{\citenamefont {Motzoi}\ and\ \citenamefont
  {Wilhelm}(2013)}]{PhysRevA.88.062318}%
  \BibitemOpen
  \bibfield  {author} {\bibinfo {author} {\bibfnamefont {F.}~\bibnamefont
  {Motzoi}}\ and\ \bibinfo {author} {\bibfnamefont {F.~K.}\ \bibnamefont
  {Wilhelm}},\ }\href {https://doi.org/10.1103/PhysRevA.88.062318} {\bibfield
  {journal} {\bibinfo  {journal} {Phys. Rev. A}\ }\textbf {\bibinfo {volume}
  {88}},\ \bibinfo {pages} {062318} (\bibinfo {year} {2013})}\BibitemShut
  {NoStop}%
\bibitem [{\citenamefont {Theis}\ \emph {et~al.}(2018)\citenamefont {Theis},
  \citenamefont {Motzoi}, \citenamefont {Machnes},\ and\ \citenamefont
  {Wilhelm}}]{Theis2018Counteracting}%
  \BibitemOpen
  \bibfield  {author} {\bibinfo {author} {\bibfnamefont {L.~S.}\ \bibnamefont
  {Theis}}, \bibinfo {author} {\bibfnamefont {F.}~\bibnamefont {Motzoi}},
  \bibinfo {author} {\bibfnamefont {S.}~\bibnamefont {Machnes}},\ and\ \bibinfo
  {author} {\bibfnamefont {F.~K.}\ \bibnamefont {Wilhelm}},\ }\href
  {https://doi.org/10.1209/0295-5075/123/60001} {\bibfield  {journal} {\bibinfo
   {journal} {EPL (Europhysics Letters)}\ }\textbf {\bibinfo {volume} {123}},\
  \bibinfo {pages} {60001} (\bibinfo {year} {2018})}\BibitemShut {NoStop}%
\bibitem [{\citenamefont {Gambetta}\ \emph {et~al.}(2011)\citenamefont
  {Gambetta}, \citenamefont {Motzoi}, \citenamefont {Merkel},\ and\
  \citenamefont {Wilhelm}}]{PhysRevA.83.012308}%
  \BibitemOpen
  \bibfield  {author} {\bibinfo {author} {\bibfnamefont {J.~M.}\ \bibnamefont
  {Gambetta}}, \bibinfo {author} {\bibfnamefont {F.}~\bibnamefont {Motzoi}},
  \bibinfo {author} {\bibfnamefont {S.~T.}\ \bibnamefont {Merkel}},\ and\
  \bibinfo {author} {\bibfnamefont {F.~K.}\ \bibnamefont {Wilhelm}},\ }\href
  {https://doi.org/10.1103/PhysRevA.83.012308} {\bibfield  {journal} {\bibinfo
  {journal} {Phys. Rev. A}\ }\textbf {\bibinfo {volume} {83}},\ \bibinfo
  {pages} {012308} (\bibinfo {year} {2011})}\BibitemShut {NoStop}%
\bibitem [{\citenamefont {Li}\ \emph {et~al.}(2024)\citenamefont {Li},
  \citenamefont {Calarco},\ and\ \citenamefont {Motzoi}}]{Li2024Experimental}%
  \BibitemOpen
  \bibfield  {author} {\bibinfo {author} {\bibfnamefont {B.}~\bibnamefont
  {Li}}, \bibinfo {author} {\bibfnamefont {T.}~\bibnamefont {Calarco}},\ and\
  \bibinfo {author} {\bibfnamefont {F.}~\bibnamefont {Motzoi}},\ }\href
  {https://doi.org/10.1038/s41534-024-00863-4} {\bibfield  {journal} {\bibinfo
  {journal} {npj Quantum Information}\ }\textbf {\bibinfo {volume} {10}},\
  \bibinfo {pages} {1} (\bibinfo {year} {2024})}\BibitemShut {NoStop}%
\bibitem [{\citenamefont {Li}\ \emph {et~al.}(2025)\citenamefont {Li},
  \citenamefont {C\'ardenas-L\'opez}, \citenamefont {Lupascu},\ and\
  \citenamefont {Motzoi}}]{li2024universal}%
  \BibitemOpen
  \bibfield  {author} {\bibinfo {author} {\bibfnamefont {B.}~\bibnamefont
  {Li}}, \bibinfo {author} {\bibfnamefont {F.}~\bibnamefont
  {C\'ardenas-L\'opez}}, \bibinfo {author} {\bibfnamefont {A.}~\bibnamefont
  {Lupascu}},\ and\ \bibinfo {author} {\bibfnamefont {F.}~\bibnamefont
  {Motzoi}},\ }\href {https://doi.org/10.1103/9dxw-4c7y} {\bibfield  {journal}
  {\bibinfo  {journal} {PRX Quantum}\ }\textbf {\bibinfo {volume} {6}},\
  \bibinfo {pages} {030357} (\bibinfo {year} {2025})}\BibitemShut {NoStop}%
\bibitem [{\citenamefont {Hyypp\"a}\ \emph {et~al.}(2024)\citenamefont
  {Hyypp\"a}, \citenamefont {Veps\"al\"ainen}, \citenamefont
  {Papi\ifmmode~\check{c}\else \v{c}\fi{}}, \citenamefont {Chan}, \citenamefont
  {Inel}, \citenamefont {Landra}, \citenamefont {Liu}, \citenamefont {Luus},
  \citenamefont {Marxer}, \citenamefont {Ockeloen-Korppi}, \citenamefont
  {Orbell}, \citenamefont {Tarasinski},\ and\ \citenamefont
  {Heinsoo}}]{PRXQuantum.5.030353}%
  \BibitemOpen
  \bibfield  {author} {\bibinfo {author} {\bibfnamefont {E.}~\bibnamefont
  {Hyypp\"a}}, \bibinfo {author} {\bibfnamefont {A.}~\bibnamefont
  {Veps\"al\"ainen}}, \bibinfo {author} {\bibfnamefont {M.}~\bibnamefont
  {Papi\ifmmode~\check{c}\else \v{c}\fi{}}}, \bibinfo {author} {\bibfnamefont
  {C.~F.}\ \bibnamefont {Chan}}, \bibinfo {author} {\bibfnamefont
  {S.}~\bibnamefont {Inel}}, \bibinfo {author} {\bibfnamefont {A.}~\bibnamefont
  {Landra}}, \bibinfo {author} {\bibfnamefont {W.}~\bibnamefont {Liu}},
  \bibinfo {author} {\bibfnamefont {J.}~\bibnamefont {Luus}}, \bibinfo {author}
  {\bibfnamefont {F.}~\bibnamefont {Marxer}}, \bibinfo {author} {\bibfnamefont
  {C.}~\bibnamefont {Ockeloen-Korppi}}, \bibinfo {author} {\bibfnamefont
  {S.}~\bibnamefont {Orbell}}, \bibinfo {author} {\bibfnamefont
  {B.}~\bibnamefont {Tarasinski}},\ and\ \bibinfo {author} {\bibfnamefont
  {J.}~\bibnamefont {Heinsoo}},\ }\href
  {https://doi.org/10.1103/PRXQuantum.5.030353} {\bibfield  {journal} {\bibinfo
   {journal} {PRX Quantum}\ }\textbf {\bibinfo {volume} {5}},\ \bibinfo {pages}
  {030353} (\bibinfo {year} {2024})}\BibitemShut {NoStop}%
\bibitem [{\citenamefont {Chiaro}\ and\ \citenamefont
  {Zhang}(2025)}]{Chiaro2025}%
  \BibitemOpen
  \bibfield  {author} {\bibinfo {author} {\bibfnamefont {B.}~\bibnamefont
  {Chiaro}}\ and\ \bibinfo {author} {\bibfnamefont {Y.}~\bibnamefont {Zhang}},\
  }\href {https://doi.org/10.1103/4kz9-w97h} {\bibfield  {journal} {\bibinfo
  {journal} {Phys. Rev. Lett.}\ }\textbf {\bibinfo {volume} {135}},\ \bibinfo
  {pages} {130601} (\bibinfo {year} {2025})}\BibitemShut {NoStop}%
\bibitem [{\citenamefont {Khani}\ \emph {et~al.}(2009)\citenamefont {Khani},
  \citenamefont {Gambetta}, \citenamefont {Motzoi},\ and\ \citenamefont
  {Wilhelm}}]{Khani2009Optimal}%
  \BibitemOpen
  \bibfield  {author} {\bibinfo {author} {\bibfnamefont {B.}~\bibnamefont
  {Khani}}, \bibinfo {author} {\bibfnamefont {J.~M.}\ \bibnamefont {Gambetta}},
  \bibinfo {author} {\bibfnamefont {F.}~\bibnamefont {Motzoi}},\ and\ \bibinfo
  {author} {\bibfnamefont {F.~K.}\ \bibnamefont {Wilhelm}},\ }\href
  {https://doi.org/10.1088/0031-8949/2009/T137/014021} {\bibfield  {journal}
  {\bibinfo  {journal} {Physica Scripta}\ }\textbf {\bibinfo {volume} {2009}},\
  \bibinfo {pages} {014021} (\bibinfo {year} {2009})}\BibitemShut {NoStop}%
\bibitem [{\citenamefont {Li}\ \emph {et~al.}(2022)\citenamefont {Li},
  \citenamefont {Calarco},\ and\ \citenamefont
  {Motzoi}}]{Li2022Nonperturbative}%
  \BibitemOpen
  \bibfield  {author} {\bibinfo {author} {\bibfnamefont {B.}~\bibnamefont
  {Li}}, \bibinfo {author} {\bibfnamefont {T.}~\bibnamefont {Calarco}},\ and\
  \bibinfo {author} {\bibfnamefont {F.}~\bibnamefont {Motzoi}},\ }\href
  {https://doi.org/10.1103/PRXQuantum.3.030313} {\bibfield  {journal} {\bibinfo
   {journal} {PRX Quantum}\ }\textbf {\bibinfo {volume} {3}},\ \bibinfo {pages}
  {030313} (\bibinfo {year} {2022})}\BibitemShut {NoStop}%
\bibitem [{\citenamefont {Schrieffer}\ and\ \citenamefont
  {Wolff}(1966)}]{PhysRev.149.491}%
  \BibitemOpen
  \bibfield  {author} {\bibinfo {author} {\bibfnamefont {J.~R.}\ \bibnamefont
  {Schrieffer}}\ and\ \bibinfo {author} {\bibfnamefont {P.~A.}\ \bibnamefont
  {Wolff}},\ }\href {https://doi.org/10.1103/PhysRev.149.491} {\bibfield
  {journal} {\bibinfo  {journal} {Phys. Rev.}\ }\textbf {\bibinfo {volume}
  {149}},\ \bibinfo {pages} {491} (\bibinfo {year} {1966})}\BibitemShut
  {NoStop}%
\bibitem [{\citenamefont {Ib{\'a}{\~n}ez}\ \emph {et~al.}(2012)\citenamefont
  {Ib{\'a}{\~n}ez}, \citenamefont {Chen}, \citenamefont {Torrontegui},
  \citenamefont {Muga},\ and\ \citenamefont {Ruschhaupt}}]{ibanez2012multiple}%
  \BibitemOpen
  \bibfield  {author} {\bibinfo {author} {\bibfnamefont {S.}~\bibnamefont
  {Ib{\'a}{\~n}ez}}, \bibinfo {author} {\bibfnamefont {X.}~\bibnamefont
  {Chen}}, \bibinfo {author} {\bibfnamefont {E.}~\bibnamefont {Torrontegui}},
  \bibinfo {author} {\bibfnamefont {J.~G.}\ \bibnamefont {Muga}},\ and\
  \bibinfo {author} {\bibfnamefont {A.}~\bibnamefont {Ruschhaupt}},\
  }\href@noop {} {\bibfield  {journal} {\bibinfo  {journal} {Physical review
  letters}\ }\textbf {\bibinfo {volume} {109}},\ \bibinfo {pages} {100403}
  (\bibinfo {year} {2012})}\BibitemShut {NoStop}%
\bibitem [{\citenamefont {Pedersen}\ \emph {et~al.}(2007)\citenamefont
  {Pedersen}, \citenamefont {M{\o}ller},\ and\ \citenamefont
  {M{\o}lmer}}]{Pedersen2007Fidelity}%
  \BibitemOpen
  \bibfield  {author} {\bibinfo {author} {\bibfnamefont {L.~H.}\ \bibnamefont
  {Pedersen}}, \bibinfo {author} {\bibfnamefont {N.~M.}\ \bibnamefont
  {M{\o}ller}},\ and\ \bibinfo {author} {\bibfnamefont {K.}~\bibnamefont
  {M{\o}lmer}},\ }\href {https://doi.org/10.1016/j.physleta.2007.02.069}
  {\bibfield  {journal} {\bibinfo  {journal} {Physics Letters A}\ }\textbf
  {\bibinfo {volume} {367}},\ \bibinfo {pages} {47} (\bibinfo {year}
  {2007})}\BibitemShut {NoStop}%
\bibitem [{\citenamefont {Wang}\ \emph {et~al.}(2025)\citenamefont {Wang},
  \citenamefont {Feng}, \citenamefont {Zhang}, \citenamefont {Ding},
  \citenamefont {Li}, \citenamefont {Motzoi}, \citenamefont {Gao},
  \citenamefont {Xu}, \citenamefont {Yang}, \citenamefont {Nuerbolati},
  \citenamefont {Yu}, \citenamefont {Sun},\ and\ \citenamefont
  {Yan}}]{Wang2025Suppressinga}%
  \BibitemOpen
  \bibfield  {author} {\bibinfo {author} {\bibfnamefont {R.}~\bibnamefont
  {Wang}}, \bibinfo {author} {\bibfnamefont {Y.}~\bibnamefont {Feng}}, \bibinfo
  {author} {\bibfnamefont {Y.}~\bibnamefont {Zhang}}, \bibinfo {author}
  {\bibfnamefont {J.}~\bibnamefont {Ding}}, \bibinfo {author} {\bibfnamefont
  {B.}~\bibnamefont {Li}}, \bibinfo {author} {\bibfnamefont {F.}~\bibnamefont
  {Motzoi}}, \bibinfo {author} {\bibfnamefont {Y.}~\bibnamefont {Gao}},
  \bibinfo {author} {\bibfnamefont {H.}~\bibnamefont {Xu}}, \bibinfo {author}
  {\bibfnamefont {Z.}~\bibnamefont {Yang}}, \bibinfo {author} {\bibfnamefont
  {W.}~\bibnamefont {Nuerbolati}}, \bibinfo {author} {\bibfnamefont
  {H.}~\bibnamefont {Yu}}, \bibinfo {author} {\bibfnamefont {W.}~\bibnamefont
  {Sun}},\ and\ \bibinfo {author} {\bibfnamefont {F.}~\bibnamefont {Yan}},\
  }\href {https://doi.org/10.1103/h4xf-vq2l} {\bibfield  {journal} {\bibinfo
  {journal} {Physical Review Letters}\ }\textbf {\bibinfo {volume} {135}},\
  \bibinfo {pages} {160804} (\bibinfo {year} {2025})}\BibitemShut {NoStop}%
\bibitem [{\citenamefont {Marxer}\ \emph {et~al.}(2025)\citenamefont {Marxer},
  \citenamefont {Mrożek}, \citenamefont {Andersson}, \citenamefont
  {Abdurakhimov}, \citenamefont {Adam}, \citenamefont {Bergholm}, \citenamefont
  {Beriwal}, \citenamefont {Chan}, \citenamefont {Dahl}, \citenamefont {Das},
  \citenamefont {Deppe}, \citenamefont {Fedorets}, \citenamefont {Gao},
  \citenamefont {Frieiro}, \citenamefont {Gusenkova}, \citenamefont {Guthrie},
  \citenamefont {Hiltunen}, \citenamefont {Hsu}, \citenamefont {Hyyppä},
  \citenamefont {Ikonen}, \citenamefont {Inel}, \citenamefont {Jolin},
  \citenamefont {Karis}, \citenamefont {Kim}, \citenamefont {Kindel},
  \citenamefont {Komlev}, \citenamefont {Koistinen}, \citenamefont
  {Kokkoniemi}, \citenamefont {Kumar}, \citenamefont {Ku}, \citenamefont
  {Lamprich}, \citenamefont {Laine}, \citenamefont {Landra}, \citenamefont
  {Lee}, \citenamefont {Lethif}, \citenamefont {Liebermann}, \citenamefont
  {Liu}, \citenamefont {Mitra}, \citenamefont {Mylläri}, \citenamefont
  {Ockeloen-Korppi}, \citenamefont {Orell}, \citenamefont {Plyshch},
  \citenamefont {Räbinä}, \citenamefont {Rebello}, \citenamefont {Renger},
  \citenamefont {Reentilä}, \citenamefont {Ritvas}, \citenamefont {Saarinen},
  \citenamefont {Salmenkivi}, \citenamefont {Sarsby}, \citenamefont
  {Savytskyi}, \citenamefont {Selinmaa}, \citenamefont {Steggles},
  \citenamefont {Takala}, \citenamefont {Takmakov}, \citenamefont {Tarasinski},
  \citenamefont {Tuorila}, \citenamefont {Välimaa}, \citenamefont {Verjauw},
  \citenamefont {Wesdorp}, \citenamefont {Wurz}, \citenamefont {Qiu},
  \citenamefont {Zhu}, \citenamefont {Hassel}, \citenamefont {Heinsoo},
  \citenamefont {Geresdi},\ and\ \citenamefont
  {Vepsäläinen}}]{marxer2025999fidelitysinglequbitgates}%
  \BibitemOpen
  \bibfield  {author} {\bibinfo {author} {\bibfnamefont {F.}~\bibnamefont
  {Marxer}}, \bibinfo {author} {\bibfnamefont {J.}~\bibnamefont {Mrożek}},
  \bibinfo {author} {\bibfnamefont {J.}~\bibnamefont {Andersson}}, \bibinfo
  {author} {\bibfnamefont {L.}~\bibnamefont {Abdurakhimov}}, \bibinfo {author}
  {\bibfnamefont {J.}~\bibnamefont {Adam}}, \bibinfo {author} {\bibfnamefont
  {V.}~\bibnamefont {Bergholm}}, \bibinfo {author} {\bibfnamefont
  {R.}~\bibnamefont {Beriwal}}, \bibinfo {author} {\bibfnamefont {C.~F.}\
  \bibnamefont {Chan}}, \bibinfo {author} {\bibfnamefont {S.}~\bibnamefont
  {Dahl}}, \bibinfo {author} {\bibfnamefont {S.~R.}\ \bibnamefont {Das}},
  \bibinfo {author} {\bibfnamefont {F.}~\bibnamefont {Deppe}}, \bibinfo
  {author} {\bibfnamefont {O.}~\bibnamefont {Fedorets}}, \bibinfo {author}
  {\bibfnamefont {Z.}~\bibnamefont {Gao}}, \bibinfo {author} {\bibfnamefont
  {A.~G.}\ \bibnamefont {Frieiro}}, \bibinfo {author} {\bibfnamefont
  {D.}~\bibnamefont {Gusenkova}}, \bibinfo {author} {\bibfnamefont
  {A.}~\bibnamefont {Guthrie}}, \bibinfo {author} {\bibfnamefont
  {T.}~\bibnamefont {Hiltunen}}, \bibinfo {author} {\bibfnamefont
  {H.}~\bibnamefont {Hsu}}, \bibinfo {author} {\bibfnamefont {E.}~\bibnamefont
  {Hyyppä}}, \bibinfo {author} {\bibfnamefont {J.}~\bibnamefont {Ikonen}},
  \bibinfo {author} {\bibfnamefont {S.}~\bibnamefont {Inel}}, \bibinfo {author}
  {\bibfnamefont {S.~W.}\ \bibnamefont {Jolin}}, \bibinfo {author}
  {\bibfnamefont {A.}~\bibnamefont {Karis}}, \bibinfo {author} {\bibfnamefont
  {S.-G.}\ \bibnamefont {Kim}}, \bibinfo {author} {\bibfnamefont
  {W.}~\bibnamefont {Kindel}}, \bibinfo {author} {\bibfnamefont
  {A.}~\bibnamefont {Komlev}}, \bibinfo {author} {\bibfnamefont
  {M.}~\bibnamefont {Koistinen}}, \bibinfo {author} {\bibfnamefont
  {R.}~\bibnamefont {Kokkoniemi}}, \bibinfo {author} {\bibfnamefont
  {S.}~\bibnamefont {Kumar}}, \bibinfo {author} {\bibfnamefont {H.-S.}\
  \bibnamefont {Ku}}, \bibinfo {author} {\bibfnamefont {J.}~\bibnamefont
  {Lamprich}}, \bibinfo {author} {\bibfnamefont {S.}~\bibnamefont {Laine}},
  \bibinfo {author} {\bibfnamefont {A.}~\bibnamefont {Landra}}, \bibinfo
  {author} {\bibfnamefont {L.-H.}\ \bibnamefont {Lee}}, \bibinfo {author}
  {\bibfnamefont {N.}~\bibnamefont {Lethif}}, \bibinfo {author} {\bibfnamefont
  {P.}~\bibnamefont {Liebermann}}, \bibinfo {author} {\bibfnamefont
  {W.}~\bibnamefont {Liu}}, \bibinfo {author} {\bibfnamefont {K.}~\bibnamefont
  {Mitra}}, \bibinfo {author} {\bibfnamefont {T.}~\bibnamefont {Mylläri}},
  \bibinfo {author} {\bibfnamefont {C.}~\bibnamefont {Ockeloen-Korppi}},
  \bibinfo {author} {\bibfnamefont {T.}~\bibnamefont {Orell}}, \bibinfo
  {author} {\bibfnamefont {A.}~\bibnamefont {Plyshch}}, \bibinfo {author}
  {\bibfnamefont {J.}~\bibnamefont {Räbinä}}, \bibinfo {author}
  {\bibfnamefont {A.}~\bibnamefont {Rebello}}, \bibinfo {author} {\bibfnamefont
  {M.}~\bibnamefont {Renger}}, \bibinfo {author} {\bibfnamefont
  {O.}~\bibnamefont {Reentilä}}, \bibinfo {author} {\bibfnamefont
  {J.}~\bibnamefont {Ritvas}}, \bibinfo {author} {\bibfnamefont
  {S.}~\bibnamefont {Saarinen}}, \bibinfo {author} {\bibfnamefont
  {O.}~\bibnamefont {Salmenkivi}}, \bibinfo {author} {\bibfnamefont
  {M.}~\bibnamefont {Sarsby}}, \bibinfo {author} {\bibfnamefont
  {M.}~\bibnamefont {Savytskyi}}, \bibinfo {author} {\bibfnamefont
  {V.}~\bibnamefont {Selinmaa}}, \bibinfo {author} {\bibfnamefont
  {M.}~\bibnamefont {Steggles}}, \bibinfo {author} {\bibfnamefont
  {E.}~\bibnamefont {Takala}}, \bibinfo {author} {\bibfnamefont
  {I.}~\bibnamefont {Takmakov}}, \bibinfo {author} {\bibfnamefont
  {B.}~\bibnamefont {Tarasinski}}, \bibinfo {author} {\bibfnamefont
  {J.}~\bibnamefont {Tuorila}}, \bibinfo {author} {\bibfnamefont
  {A.}~\bibnamefont {Välimaa}}, \bibinfo {author} {\bibfnamefont
  {J.}~\bibnamefont {Verjauw}}, \bibinfo {author} {\bibfnamefont
  {J.}~\bibnamefont {Wesdorp}}, \bibinfo {author} {\bibfnamefont
  {N.}~\bibnamefont {Wurz}}, \bibinfo {author} {\bibfnamefont {W.}~\bibnamefont
  {Qiu}}, \bibinfo {author} {\bibfnamefont {L.}~\bibnamefont {Zhu}}, \bibinfo
  {author} {\bibfnamefont {J.}~\bibnamefont {Hassel}}, \bibinfo {author}
  {\bibfnamefont {J.}~\bibnamefont {Heinsoo}}, \bibinfo {author} {\bibfnamefont
  {A.}~\bibnamefont {Geresdi}},\ and\ \bibinfo {author} {\bibfnamefont
  {A.}~\bibnamefont {Vepsäläinen}},\ }\href
  {https://arxiv.org/abs/2508.16437} {\bibinfo {title} {Above 99.9% fidelity
  single-qubit gates, two-qubit gates, and readout in a single superconducting
  quantum device}} (\bibinfo {year} {2025}),\ \Eprint
  {https://arxiv.org/abs/2508.16437} {arXiv:2508.16437 [quant-ph]} \BibitemShut
  {NoStop}%
\bibitem [{\citenamefont {Burnett}\ \emph {et~al.}(2019)\citenamefont
  {Burnett}, \citenamefont {Bengtsson}, \citenamefont {Scigliuzzo},
  \citenamefont {Niepce}, \citenamefont {Kudra}, \citenamefont {Delsing},\ and\
  \citenamefont {Bylander}}]{Burnettdriftt1t2}%
  \BibitemOpen
  \bibfield  {author} {\bibinfo {author} {\bibfnamefont {J.~J.}\ \bibnamefont
  {Burnett}}, \bibinfo {author} {\bibfnamefont {A.}~\bibnamefont {Bengtsson}},
  \bibinfo {author} {\bibfnamefont {M.}~\bibnamefont {Scigliuzzo}}, \bibinfo
  {author} {\bibfnamefont {D.}~\bibnamefont {Niepce}}, \bibinfo {author}
  {\bibfnamefont {M.}~\bibnamefont {Kudra}}, \bibinfo {author} {\bibfnamefont
  {P.}~\bibnamefont {Delsing}},\ and\ \bibinfo {author} {\bibfnamefont
  {J.}~\bibnamefont {Bylander}},\ }\href
  {https://doi.org/10.1038/s41534-019-0168-5} {\bibfield  {journal} {\bibinfo
  {journal} {npj Quantum Information}\ }\textbf {\bibinfo {volume} {5}},\
  \bibinfo {pages} {54} (\bibinfo {year} {2019})}\BibitemShut {NoStop}%
\bibitem [{\citenamefont {de~Graaf}\ \emph {et~al.}(2020)\citenamefont
  {de~Graaf}, \citenamefont {Faoro}, \citenamefont {Ioffe}, \citenamefont
  {Mahashabde}, \citenamefont {Burnett}, \citenamefont {Lindström},
  \citenamefont {Kubatkin}, \citenamefont {Danilov},\ and\ \citenamefont
  {Tzalenchuk}}]{graaffcali}%
  \BibitemOpen
  \bibfield  {author} {\bibinfo {author} {\bibfnamefont {S.~E.}\ \bibnamefont
  {de~Graaf}}, \bibinfo {author} {\bibfnamefont {L.}~\bibnamefont {Faoro}},
  \bibinfo {author} {\bibfnamefont {L.~B.}\ \bibnamefont {Ioffe}}, \bibinfo
  {author} {\bibfnamefont {S.}~\bibnamefont {Mahashabde}}, \bibinfo {author}
  {\bibfnamefont {J.~J.}\ \bibnamefont {Burnett}}, \bibinfo {author}
  {\bibfnamefont {T.}~\bibnamefont {Lindström}}, \bibinfo {author}
  {\bibfnamefont {S.~E.}\ \bibnamefont {Kubatkin}}, \bibinfo {author}
  {\bibfnamefont {A.~V.}\ \bibnamefont {Danilov}},\ and\ \bibinfo {author}
  {\bibfnamefont {A.~Y.}\ \bibnamefont {Tzalenchuk}},\ }\href
  {https://doi.org/10.1126/sciadv.abc5055} {\bibfield  {journal} {\bibinfo
  {journal} {Science Advances}\ }\textbf {\bibinfo {volume} {6}},\ \bibinfo
  {pages} {eabc5055} (\bibinfo {year} {2020})},\ \Eprint
  {https://arxiv.org/abs/https://www.science.org/doi/pdf/10.1126/sciadv.abc5055}
  {https://www.science.org/doi/pdf/10.1126/sciadv.abc5055} \BibitemShut
  {NoStop}%
\bibitem [{\citenamefont {Wu}\ \emph {et~al.}(2024)\citenamefont {Wu},
  \citenamefont {Lin}, \citenamefont {Xie}, \citenamefont {Guo}, \citenamefont
  {Huang}, \citenamefont {Zhang}, \citenamefont {Zhou}, \citenamefont {Sun},
  \citenamefont {Zhang}, \citenamefont {Guo}, \citenamefont {Linpeng},
  \citenamefont {Liu}, \citenamefont {Liu}, \citenamefont {Ren}, \citenamefont
  {Tao}, \citenamefont {Jiang}, \citenamefont {Chu}, \citenamefont {Niu},
  \citenamefont {Zhong},\ and\ \citenamefont {Yu}}]{wudrift}%
  \BibitemOpen
  \bibfield  {author} {\bibinfo {author} {\bibfnamefont {N.}~\bibnamefont
  {Wu}}, \bibinfo {author} {\bibfnamefont {J.}~\bibnamefont {Lin}}, \bibinfo
  {author} {\bibfnamefont {C.}~\bibnamefont {Xie}}, \bibinfo {author}
  {\bibfnamefont {Z.}~\bibnamefont {Guo}}, \bibinfo {author} {\bibfnamefont
  {W.}~\bibnamefont {Huang}}, \bibinfo {author} {\bibfnamefont
  {L.}~\bibnamefont {Zhang}}, \bibinfo {author} {\bibfnamefont
  {Y.}~\bibnamefont {Zhou}}, \bibinfo {author} {\bibfnamefont {X.}~\bibnamefont
  {Sun}}, \bibinfo {author} {\bibfnamefont {J.}~\bibnamefont {Zhang}}, \bibinfo
  {author} {\bibfnamefont {W.}~\bibnamefont {Guo}}, \bibinfo {author}
  {\bibfnamefont {X.}~\bibnamefont {Linpeng}}, \bibinfo {author} {\bibfnamefont
  {S.}~\bibnamefont {Liu}}, \bibinfo {author} {\bibfnamefont {Y.}~\bibnamefont
  {Liu}}, \bibinfo {author} {\bibfnamefont {W.}~\bibnamefont {Ren}}, \bibinfo
  {author} {\bibfnamefont {Z.}~\bibnamefont {Tao}}, \bibinfo {author}
  {\bibfnamefont {J.}~\bibnamefont {Jiang}}, \bibinfo {author} {\bibfnamefont
  {J.}~\bibnamefont {Chu}}, \bibinfo {author} {\bibfnamefont {J.}~\bibnamefont
  {Niu}}, \bibinfo {author} {\bibfnamefont {Y.}~\bibnamefont {Zhong}},\ and\
  \bibinfo {author} {\bibfnamefont {D.}~\bibnamefont {Yu}},\ }\href
  {https://doi.org/10.1063/5.0234579} {\bibfield  {journal} {\bibinfo
  {journal} {Applied Physics Letters}\ }\textbf {\bibinfo {volume} {125}},\
  \bibinfo {pages} {204003} (\bibinfo {year} {2024})}\BibitemShut {NoStop}%
\bibitem [{\citenamefont {Gustavsson}\ \emph {et~al.}(2013)\citenamefont
  {Gustavsson}, \citenamefont {Zwier}, \citenamefont {Bylander}, \citenamefont
  {Yan}, \citenamefont {Yoshihara}, \citenamefont {Nakamura}, \citenamefont
  {Orlando},\ and\ \citenamefont {Oliver}}]{PhysRevLett.110.040502}%
  \BibitemOpen
  \bibfield  {author} {\bibinfo {author} {\bibfnamefont {S.}~\bibnamefont
  {Gustavsson}}, \bibinfo {author} {\bibfnamefont {O.}~\bibnamefont {Zwier}},
  \bibinfo {author} {\bibfnamefont {J.}~\bibnamefont {Bylander}}, \bibinfo
  {author} {\bibfnamefont {F.}~\bibnamefont {Yan}}, \bibinfo {author}
  {\bibfnamefont {F.}~\bibnamefont {Yoshihara}}, \bibinfo {author}
  {\bibfnamefont {Y.}~\bibnamefont {Nakamura}}, \bibinfo {author}
  {\bibfnamefont {T.~P.}\ \bibnamefont {Orlando}},\ and\ \bibinfo {author}
  {\bibfnamefont {W.~D.}\ \bibnamefont {Oliver}},\ }\href
  {https://doi.org/10.1103/PhysRevLett.110.040502} {\bibfield  {journal}
  {\bibinfo  {journal} {Phys. Rev. Lett.}\ }\textbf {\bibinfo {volume} {110}},\
  \bibinfo {pages} {040502} (\bibinfo {year} {2013})}\BibitemShut {NoStop}%
\bibitem [{\citenamefont {Guo}\ \emph {et~al.}(2024)\citenamefont {Guo},
  \citenamefont {Duan}, \citenamefont {Zhang}, \citenamefont {Yang},
  \citenamefont {Zhang}, \citenamefont {Du}, \citenamefont {Zhang},
  \citenamefont {Tao}, \citenamefont {Wang}, \citenamefont {Jia}, \citenamefont
  {Chen},\ and\ \citenamefont {Guo}}]{PhysRevApplied.21.064060}%
  \BibitemOpen
  \bibfield  {author} {\bibinfo {author} {\bibfnamefont {L.-L.}\ \bibnamefont
  {Guo}}, \bibinfo {author} {\bibfnamefont {P.}~\bibnamefont {Duan}}, \bibinfo
  {author} {\bibfnamefont {S.}~\bibnamefont {Zhang}}, \bibinfo {author}
  {\bibfnamefont {X.-X.}\ \bibnamefont {Yang}}, \bibinfo {author}
  {\bibfnamefont {C.}~\bibnamefont {Zhang}}, \bibinfo {author} {\bibfnamefont
  {L.}~\bibnamefont {Du}}, \bibinfo {author} {\bibfnamefont {H.-F.}\
  \bibnamefont {Zhang}}, \bibinfo {author} {\bibfnamefont {H.-R.}\ \bibnamefont
  {Tao}}, \bibinfo {author} {\bibfnamefont {T.-L.}\ \bibnamefont {Wang}},
  \bibinfo {author} {\bibfnamefont {Z.-L.}\ \bibnamefont {Jia}}, \bibinfo
  {author} {\bibfnamefont {Z.-Y.}\ \bibnamefont {Chen}},\ and\ \bibinfo
  {author} {\bibfnamefont {G.-P.}\ \bibnamefont {Guo}},\ }\href
  {https://doi.org/10.1103/PhysRevApplied.21.064060} {\bibfield  {journal}
  {\bibinfo  {journal} {Phys. Rev. Appl.}\ }\textbf {\bibinfo {volume} {21}},\
  \bibinfo {pages} {064060} (\bibinfo {year} {2024})}\BibitemShut {NoStop}%
\bibitem [{\citenamefont {Warren}(1984)}]{Warrenmagnus}%
  \BibitemOpen
  \bibfield  {author} {\bibinfo {author} {\bibfnamefont {W.~S.}\ \bibnamefont
  {Warren}},\ }\href {https://doi.org/10.1063/1.447644} {\bibfield  {journal}
  {\bibinfo  {journal} {The Journal of Chemical Physics}\ }\textbf {\bibinfo
  {volume} {81}},\ \bibinfo {pages} {5437} (\bibinfo {year}
  {1984})}\BibitemShut {NoStop}%
\bibitem [{\citenamefont {Haeberlen}(1976)}]{Haeberlen1976}%
  \BibitemOpen
  \bibfield  {author} {\bibinfo {author} {\bibfnamefont {U.}~\bibnamefont
  {Haeberlen}},\ }\href@noop {} {\emph {\bibinfo {title} {High Resolution NMR
  in Solids: Selective Averaging}}},\ Advances in Magnetic Resonance,
  Supplement 1\ (\bibinfo  {publisher} {Academic Press},\ \bibinfo {address}
  {New York, NY},\ \bibinfo {year} {1976})\BibitemShut {NoStop}%
\bibitem [{\citenamefont {Brinkmann}(2016)}]{Brinkmann2016}%
  \BibitemOpen
  \bibfield  {author} {\bibinfo {author} {\bibfnamefont {A.}~\bibnamefont
  {Brinkmann}},\ }\href {https://doi.org/10.1002/cmr.a.21414} {\bibfield
  {journal} {\bibinfo  {journal} {Concepts in Magnetic Resonance Part A}\
  }\textbf {\bibinfo {volume} {45A}},\ \bibinfo {pages} {e21414} (\bibinfo
  {year} {2016})}\BibitemShut {NoStop}%
\bibitem [{\citenamefont {Blanes}\ \emph {et~al.}(2009)\citenamefont {Blanes},
  \citenamefont {Casas}, \citenamefont {Oteo},\ and\ \citenamefont
  {Ros}}]{Blanes_2009}%
  \BibitemOpen
  \bibfield  {author} {\bibinfo {author} {\bibfnamefont {S.}~\bibnamefont
  {Blanes}}, \bibinfo {author} {\bibfnamefont {F.}~\bibnamefont {Casas}},
  \bibinfo {author} {\bibfnamefont {J.}~\bibnamefont {Oteo}},\ and\ \bibinfo
  {author} {\bibfnamefont {J.}~\bibnamefont {Ros}},\ }\href
  {https://doi.org/10.1016/j.physrep.2008.11.001} {\bibfield  {journal}
  {\bibinfo  {journal} {Physics Reports}\ }\textbf {\bibinfo {volume} {470}},\
  \bibinfo {pages} {151–238} (\bibinfo {year} {2009})}\BibitemShut {NoStop}%
\bibitem [{\citenamefont {Kirchhoff}\ \emph
  {et~al.}(2025{\natexlab{a}})\citenamefont {Kirchhoff}, \citenamefont
  {Wilhelm},\ and\ \citenamefont {Motzoi}}]{PRXQuantum.6.010328}%
  \BibitemOpen
  \bibfield  {author} {\bibinfo {author} {\bibfnamefont {S.}~\bibnamefont
  {Kirchhoff}}, \bibinfo {author} {\bibfnamefont {F.~K.}\ \bibnamefont
  {Wilhelm}},\ and\ \bibinfo {author} {\bibfnamefont {F.}~\bibnamefont
  {Motzoi}},\ }\href {https://doi.org/10.1103/PRXQuantum.6.010328} {\bibfield
  {journal} {\bibinfo  {journal} {PRX Quantum}\ }\textbf {\bibinfo {volume}
  {6}},\ \bibinfo {pages} {010328} (\bibinfo {year}
  {2025}{\natexlab{a}})}\BibitemShut {NoStop}%
\bibitem [{\citenamefont {Gao}\ \emph {et~al.}(2025)\citenamefont {Gao},
  \citenamefont {Galicia}, \citenamefont {Jesus}, \citenamefont {Liu},
  \citenamefont {Haddad}, \citenamefont {Volkov}, \citenamefont {Guimarães},
  \citenamefont {Bhardwaj}, \citenamefont {Jerger}, \citenamefont {Neis},
  \citenamefont {Li}, \citenamefont {Cárdenas-López}, \citenamefont {Motzoi},
  \citenamefont {Bushev},\ and\ \citenamefont {Barends}}]{GAO2025}%
  \BibitemOpen
  \bibfield  {author} {\bibinfo {author} {\bibfnamefont {Y.}~\bibnamefont
  {Gao}}, \bibinfo {author} {\bibfnamefont {A.}~\bibnamefont {Galicia}},
  \bibinfo {author} {\bibfnamefont {J.~D. D.~C.}\ \bibnamefont {Jesus}},
  \bibinfo {author} {\bibfnamefont {Y.}~\bibnamefont {Liu}}, \bibinfo {author}
  {\bibfnamefont {Y.}~\bibnamefont {Haddad}}, \bibinfo {author} {\bibfnamefont
  {D.~A.}\ \bibnamefont {Volkov}}, \bibinfo {author} {\bibfnamefont {J.~R.}\
  \bibnamefont {Guimarães}}, \bibinfo {author} {\bibfnamefont
  {H.}~\bibnamefont {Bhardwaj}}, \bibinfo {author} {\bibfnamefont
  {M.}~\bibnamefont {Jerger}}, \bibinfo {author} {\bibfnamefont
  {M.}~\bibnamefont {Neis}}, \bibinfo {author} {\bibfnamefont {B.}~\bibnamefont
  {Li}}, \bibinfo {author} {\bibfnamefont {F.~A.}\ \bibnamefont
  {Cárdenas-López}}, \bibinfo {author} {\bibfnamefont {F.}~\bibnamefont
  {Motzoi}}, \bibinfo {author} {\bibfnamefont {P.~A.}\ \bibnamefont {Bushev}},\
  and\ \bibinfo {author} {\bibfnamefont {R.}~\bibnamefont {Barends}},\ }\href
  {https://arxiv.org/abs/2511.22365} {\bibinfo {title} {Ultrafast single qubit
  gates through multi-photon transition removal}} (\bibinfo {year} {2025}),\
  \Eprint {https://arxiv.org/abs/2511.22365} {arXiv:2511.22365 [quant-ph]}
  \BibitemShut {NoStop}%
\bibitem [{\citenamefont {Manzano}(2020)}]{Manzano_2020}%
  \BibitemOpen
  \bibfield  {author} {\bibinfo {author} {\bibfnamefont {D.}~\bibnamefont
  {Manzano}},\ }\bibfield  {journal} {\bibinfo  {journal} {AIP Advances}\
  }\textbf {\bibinfo {volume} {10}},\ \href {https://doi.org/10.1063/1.5115323}
  {10.1063/1.5115323} (\bibinfo {year} {2020})\BibitemShut {NoStop}%
\bibitem [{\citenamefont {{IBM Quantum}}(2025)}]{IBM_Machine}%
  \BibitemOpen
  \bibfield  {author} {\bibinfo {author} {\bibnamefont {{IBM Quantum}}},\
  }\href@noop {} {\bibinfo {title} {Ibm quantum platform}},\ \bibinfo
  {howpublished}
  {\url{https://quantum.cloud.ibm.com/computers?system=ibm_pittsburgh}}
  (\bibinfo {year} {2025}),\ \bibinfo {note} {accessed: 2025-12-10}\BibitemShut
  {NoStop}%
\bibitem [{\citenamefont {Wang}\ \emph {et~al.}(2022)\citenamefont {Wang},
  \citenamefont {Li}, \citenamefont {Xu}, \citenamefont {Li}, \citenamefont
  {Wang}, \citenamefont {Yang}, \citenamefont {Mi}, \citenamefont {Liang},
  \citenamefont {Su}, \citenamefont {Yang} \emph {et~al.}}]{wang2022towards}%
  \BibitemOpen
  \bibfield  {author} {\bibinfo {author} {\bibfnamefont {C.}~\bibnamefont
  {Wang}}, \bibinfo {author} {\bibfnamefont {X.}~\bibnamefont {Li}}, \bibinfo
  {author} {\bibfnamefont {H.}~\bibnamefont {Xu}}, \bibinfo {author}
  {\bibfnamefont {Z.}~\bibnamefont {Li}}, \bibinfo {author} {\bibfnamefont
  {J.}~\bibnamefont {Wang}}, \bibinfo {author} {\bibfnamefont {Z.}~\bibnamefont
  {Yang}}, \bibinfo {author} {\bibfnamefont {Z.}~\bibnamefont {Mi}}, \bibinfo
  {author} {\bibfnamefont {X.}~\bibnamefont {Liang}}, \bibinfo {author}
  {\bibfnamefont {T.}~\bibnamefont {Su}}, \bibinfo {author} {\bibfnamefont
  {C.}~\bibnamefont {Yang}}, \emph {et~al.},\ }\href@noop {} {\bibfield
  {journal} {\bibinfo  {journal} {npj Quantum Information}\ }\textbf {\bibinfo
  {volume} {8}},\ \bibinfo {pages} {3} (\bibinfo {year} {2022})}\BibitemShut
  {NoStop}%
\bibitem [{\citenamefont {Bowdrey}\ \emph {et~al.}(2002)\citenamefont
  {Bowdrey}, \citenamefont {Oi}, \citenamefont {Short}, \citenamefont
  {Banaszek},\ and\ \citenamefont {Jones}}]{BOWDREY2002258}%
  \BibitemOpen
  \bibfield  {author} {\bibinfo {author} {\bibfnamefont {M.~D.}\ \bibnamefont
  {Bowdrey}}, \bibinfo {author} {\bibfnamefont {D.~K.}\ \bibnamefont {Oi}},
  \bibinfo {author} {\bibfnamefont {A.~J.}\ \bibnamefont {Short}}, \bibinfo
  {author} {\bibfnamefont {K.}~\bibnamefont {Banaszek}},\ and\ \bibinfo
  {author} {\bibfnamefont {J.~A.}\ \bibnamefont {Jones}},\ }\href
  {https://doi.org/https://doi.org/10.1016/S0375-9601(02)00069-5} {\bibfield
  {journal} {\bibinfo  {journal} {Physics Letters A}\ }\textbf {\bibinfo
  {volume} {294}},\ \bibinfo {pages} {258} (\bibinfo {year}
  {2002})}\BibitemShut {NoStop}%
\bibitem [{\citenamefont {Kirchhoff}\ \emph
  {et~al.}(2025{\natexlab{b}})\citenamefont {Kirchhoff}, \citenamefont
  {Wilhelm},\ and\ \citenamefont {Motzoi}}]{kirchhoff2025correction}%
  \BibitemOpen
  \bibfield  {author} {\bibinfo {author} {\bibfnamefont {S.}~\bibnamefont
  {Kirchhoff}}, \bibinfo {author} {\bibfnamefont {F.~K.}\ \bibnamefont
  {Wilhelm}},\ and\ \bibinfo {author} {\bibfnamefont {F.}~\bibnamefont
  {Motzoi}},\ }\href@noop {} {\bibfield  {journal} {\bibinfo  {journal} {PRX
  quantum}\ }\textbf {\bibinfo {volume} {6}},\ \bibinfo {pages} {010328}
  (\bibinfo {year} {2025}{\natexlab{b}})}\BibitemShut {NoStop}%
\bibitem [{\citenamefont {Jesus}\ \emph {et~al.}(2026)\citenamefont {Jesus},
  \citenamefont {Gao}, \citenamefont {Li}, \citenamefont {Barends},
  \citenamefont {C\'ardenas-L\'opez},\ and\ \citenamefont {Motzoi}}]{Repo}%
  \BibitemOpen
  \bibfield  {author} {\bibinfo {author} {\bibfnamefont {J.~D. D.~C.}\
  \bibnamefont {Jesus}}, \bibinfo {author} {\bibfnamefont {Y.}~\bibnamefont
  {Gao}}, \bibinfo {author} {\bibfnamefont {B.}~\bibnamefont {Li}}, \bibinfo
  {author} {\bibfnamefont {R.}~\bibnamefont {Barends}}, \bibinfo {author}
  {\bibfnamefont {F.~A.}\ \bibnamefont {C\'ardenas-L\'opez}},\ and\ \bibinfo
  {author} {\bibfnamefont {F.}~\bibnamefont {Motzoi}},\ }\href
  {https://doi.org/10.5281/zenodo.18744365} {\bibinfo {title} {Data produced
  for paper "analytical blueprint for 99.999\% fidelity x-gates on present
  superconducting hardware under strong driving" version 2}},\ \bibinfo
  {howpublished} {\url{https://zenodo.org/records/20341166}} (\bibinfo {year}
  {2026})\BibitemShut {NoStop}%
\end{thebibliography}%

\end{document}